\documentclass[12pt]{iopart}
\eqnobysec 
\usepackage{iopams} 
\usepackage{amsmath}  
\usepackage{amsfonts} 
\usepackage{textcomp}
\usepackage[utf8]{inputenc} 
\usepackage[T1]{fontenc}    
\usepackage{hyperref}       
\usepackage{url}            
\usepackage{booktabs}       
\usepackage{amsfonts}       
\usepackage{nicefrac}       
\usepackage{microtype}      
\usepackage{lipsum}
\usepackage[pdftex]{graphicx}
\usepackage{caption}
\usepackage{amssymb,amsmath, dsfont, mathtools, amsthm}
\usepackage{xcolor}
\usepackage{mathrsfs}
\usepackage{tensor}
\usepackage{bm}
\usepackage{todonotes}
\usepackage{cuted}
\newcommand{\ZZ}{\mathbb{Z}}
\newcommand{\CC}{\mathbb{C}}
\newcommand{\RR}{\mathbb{R}}

\newcommand{\FF}{\mathcal{F}}
\newcommand{\WW}{\mathcal{W}}

\newcommand{\OO}{\mathcal{O}}
\newcommand{\Ebar}{\overline{E}}
\newcommand{\Eabs}{|\overline{E}|}

\newcommand{\xx}{\textsf{x}}

\newcommand{\ii}{\mathrm{i}}
\newcommand{\ee}{\mathrm{e}}
\newcommand{\dd}{\mathrm{d}}

\newcommand{\DFB}{\Delta\FF_\beta}
\newcommand{\FFL}{\FF_{\mathrm{Lin}}}

\usepackage{float}
\usepackage{tikz-cd}
\usepackage{amsthm}
\usepackage{geometry}
 \usepackage{braket}

\DeclareMathOperator{\sgn}{sgn}
\DeclareMathOperator{\PolyLog}{Li}

\theoremstyle{plain}

\theoremstyle{definition}

\begin{document}

\maketitle

\title[Circular motion analogue Unruh effect
in a $2+1$ thermal bath]{Circular motion analogue Unruh effect
in a $2+1$ thermal bath: 
Robbing from the rich and giving to the poor}


\author{Cameron R D Bunney and Jorma Louko}

\address{School of Mathematical Sciences, University of Nottingham, Nottingham NG$7$ $2$RD, UK}
\ead{\mailto{cameron.bunney@nottingham.ac.uk}, \mailto{jorma.louko@nottingham.ac.uk}}

\vspace{10pt}
\begin{indented}\item[]{March 2023; revised May 2023.\footnote[2]{Published in Classical and Quantum Gravity \textbf{40}, 155001 (2023), 
doi:10.1088/1361-6382/acde3b. 
For Open Access purposes, 
this Author Accepted Manuscript is made available under CC BY public copyright.}}
\end{indented}

\begin{abstract}
The Unruh effect states that a uniformly linearly accelerated observer with proper acceleration $a$ experiences the Minkowski vacuum as a thermal state at temperature $T_U=a/(2\pi)$. An observer in uniform circular motion experiences a similar effective temperature, operationally defined in terms of excitation and de-excitation rates, and physically interpretable in terms of synchrotron radiation, but this effective temperature depends not just on the acceleration but also on the orbital speed and the excitation energy. In this paper we consider an observer in uniform circular motion when the Minkowski vacuum is replaced by an ambient thermal bath, and we address the interplay of ambient temperature, Doppler effect,  acceleration, and excitation energy. Specifically, we consider a massless scalar field in $2 + 1$ spacetime dimensions, probed by an Unruh-DeWitt detector, in a Minkowski (rather than proper) time formulation: this setting describes proposed analogue spacetime systems in which the effect may become experimentally testable, and in which an ambient temperature will necessarily be present. We establish analytic results for the observer's effective temperature in several asymptotic regions of the parameter space and provide numerical results in the interpolating regions, finding that an acceleration effect can be identified even when the Doppler effect dominates the overall magnitude of the response. We also identify parameter regimes where the observer sees a temperature lower than the ambient temperature, experiencing a cooling Unruh effect. 

\end{abstract}


\section{Introduction}
The Unruh effect \cite{Fulling,Davies1975,Unruh,Fulling:2014} is a remarkable result in quantum field theory, stating that a uniformly linearly accelerated observer with proper acceleration $a$ in Minkowski spacetime reacts to a quantum field in its Minkowski vacuum through excitations and de-excitations with the characteristics of a thermal state, at the Unruh temperature $T_U=a\hbar/(2\pi ck_B)$, proportional to the acceleration. Direct experimental verification is still, however, unconfirmed. A large hurdle to overcome is the sheer magnitude of acceleration required to reach a detectable increase in temperature. Experimental confirmation retains broad interest in its relation to the Hawking effect \cite{Hawking:1975vcx}, and the connections to the early universe quantum effects, which may originate the present-day structure of the Universe \cite{Parker:1969au,Mukhanov:2007zz}.

Phenomena similar to the Unruh effect exist also for non-linear uniform motion \cite{Letaw,Korsbakken:2004bv,Good:2020hav}, 
including uniform circular motion \cite{LetawPfautsch,Takagi,Doukas:2010wt,JIN201497,Jin:2014spa}. 
Experimental interest in the circular motion Unruh effect has a long standing \cite{BellLeinaas,Bell:1986ir,Costa:1994yx,Guimaraes:1998jf,Leinaas:1998tu,Unruh:1998gq,Lochan:2019osm}, 
in which a new angle was opened by recent proposals 
\cite{BEC1,Marino:2020uqj,Gooding,Bunney:2023ude} to utilise the analogue spacetime 
that occurs in nonrelativistic laboratory systems \cite{Unruh1981,Liberati,HeliumUniverse}. 

In analogue spacetime, circular motion enjoys two main advantages over linear acceleration. First, the experiment can remain within a finite-size laboratory for an arbitrarily long interaction time. Second, the time dilation Lorentz factor between the laboratory and the accelerated worldline remains constant in time: this allows the inclusion of the time dilation gamma-factor by appropriately scaling the energies in the theoretical analysis of the experiment, without the need to engineer a time-dependent energy scaling in a condensed matter system. Notwithstanding these advantages, a complication in circular motion is that the linear acceleration Unruh temperature formula is no longer directly applicable, and the effective temperature, operationally defined in terms of excitation and de-excitation probabilities in the accelerating system, involves also dependency on the orbital speed and the excitation energy. The underlying reason for this is that the circular motion effect does not admit a description in terms of a genuine thermal equilibrium state adapted to the motion, but has instead a physical interpretation in terms of synchrotron radiation, as reviewed in~\cite{Fulling:2014}.
A detailed comparison of the linear and circular acceleration Unruh temperatures in $2+1$ and $3+1$ spacetime dimensions is given in \cite{Biermann}. Related earlier analyses are given in \cite{Good:2020hav,Takagi,Costa:1994yx,Guimaraes:1998jf,Unruh:1998gq,Muller:1995vk,Hodgkinson}. 

In the conventional setting of the Unruh effect, 
the ambient quantum field is prepared in its Minkowski vacuum, with zero temperature --- an idealisation that no experimental test of the effect would be able to completely mimic. 
The purpose of this paper is to address the circular motion Unruh effect when the ambient quantum field is prepared in a thermal state, with a positive ambient temperature. 
Related earlier analyses are given in \cite{Costa:1994yx,Guimaraes:1998jf,Hodgkinson,Barman:2022utm}. 

We assume the circular motion to have no drift in the rest frame of the ambient heat bath. The total system is then invariant under time translations along the trajectory, and the Unruh effect will be time independent. We further specialise to a massless scalar field in $2+1$ dimensions, and we probe the field with a pointlike Unruh-DeWitt (UDW) detector \cite{Unruh,DeWitt:1980hx}. 
Finally, whereas UDW detectors normally have their transition energies defined with respect to the relativistic proper time along the detector's trajectory~\cite{Unruh,DeWitt:1980hx}, 
we define the transition energies with respect to the Minkowski time in the heat bath's rest frame. 
This setting describes proposed analogue spacetime systems in which the effect may become experimentally testable, 
and in which an ambient temperature will necessarily be present \cite{Gooding,Bunney:2023ude}.

We work in linear perturbation theory in the coupling between the detector and the field, in the limit of long interaction time but negligible back-action. We do not address finite interaction time effects \cite{Fewster} or the back-action of the detector on the field~\cite{Lin:2006jw}. 

A positive ambient temperature however brings up one new technical issue that needs to be addressed. The thermal Wightman function for a massless scalar field is well defined in spacetime dimensions $3+1$ \cite{Hodgkinson} and higher, but in $2+1$ dimensions it is infrared divergent~\cite{Takagi}. We sidestep this divergence by considering an UDW detector that couples linearly to the time derivative of the scalar field along the trajectory, 
rather than to the value of the field. The derivative-coupled detector is often employed to sidestep a similar infrared divergence that occurs for a massless field in $1+1$ spacetime dimensions already in zero temperature \cite{1991RSPSA.435..205R,Raval:1995mb,Wang:2013lex,Juarez-Aubry:2014jba,Juarez-Aubry:2018ofz,Juarez-Aubry:2021tae}. 

We first obtain a mode sum expression for the response function of the UDW detector, allowing the field to have any dispersion relation subject to mild monotonicity assumptions. In the analogue spacetime setting, this allows field frequencies that go beyond the phononic regime \cite{BEC1,Liberati}; in a fundamental relativistic spacetime setting, this allows dispersion relations that might arise from Planck scale physics \cite{Amelino-Camelia:2008aez}. 

We then specialise to the massless Klein-Gordon field. We establish analytically the asymptotic behaviour of the response function in several regimes, including the low and high ambient temperature regimes, and the corresponding asymptotic behaviour of the effective temperature, defined operationally via the detailed balance relation between excitations and de-excitations. We present numerical results for the interpolating regimes, chosen for their potential relevance for prospective experiments \cite{Gooding,Bunney:2023ude}. 

As a highlight, we show that an acceleration
effect can be identified even when the Doppler effect dominates the overall magnitude of the detector's response,as is expected to be the case in the Helium analogue spacetime system considered in~\cite{Bunney:2023ude}.
We also identify regimes in which the detector 
experiences an effective temperature that is \emph{lower\/} than the ambient temperature, so that the Unruh effect induces cooling rather than heating. 
Criteria by which the Unruh effect may be argued to induce cooling have been discussed in a variety of relativistic spacetime settings 
\cite{Brenna:2015fga,Liu:2016ihf,Garay:2016cpf,Li:2018xil,Henderson:2019uqo,Pan:2020tzf,Pan:2021nka,DeSouzaCampos:2020ddx,Barman:2021oum,Zhou:2021nyv,Robbins:2021ion,Chen:2021evr,Wu:2022rmv}. 

As a mathematical side outcome, we evaluate in closed form 
an infinite series \eqref{eq:besselsum-closed} involving squared Bessel functions. 
We have not encountered this identity in the existing literature. 

We begin in \Sref{sec: field and detector} by establishing the preliminaries for an UDW detector in uniform circular motion in $(2+1)$-dimensional Minkowski spacetime, coupled to the time derivative of a quantised real scalar field that is prepared in a thermal state. We obtain a mode sum expression for the detector's response function, recall the detailed balance definition of an effective temperature, 
reviewing its motivations and limitations,
and we investigate general conditions under which the detailed balance temperature could be expected to be lower than the ambient temperature. We also present a corresponding discussion for an inertial detector at a constant velocity with respect to the heat bath. 

\Sref{sec: limits} investigates analytically the detailed balance temperature in several limiting regimes of the parameter space, both for circular motion and for inertial motion. Interpolating numerical results are provided in \Sref{sec: numerical}. 
Separating the acceleration contribution from the Doppler contribution in the detector's response is addressed in \Sref{sec:accel-versus-speed}, by a combination of analytics and numerics. 
\Sref{sec: conclusions} presents a summary and concluding remarks. 
Proofs of technical results are deferred to two appendices.

We use units in which $c=\hbar=k_B=1$, where $c$ is the speed of light in the relativistic spacetime interpretation and the speed of sound in the analogue spacetime interpretation. 
Sans serif letters ($\xx$) denote spacetime points 
and boldface Italic letters ($\bm{k}$) denote spatial vectors. 
In asymptotic formulae, $f(x)=\OO\bigl(g(x)\bigr)$ denotes that $f(x)/g(x)$ remains bounded in the limit considered, and $f(x)=o\bigl(g(x)\bigr)$ denotes that $f(x)/g(x)$ tends to zero in the limit considered.

\section{Field and detector preliminaries}\label{sec: field and detector}

In this section we review the relevant background for an UDW detector on a circular trajectory in $(2+1)$-dimensional Minkowski spacetime, coupled to a real scalar field in a thermal state. We work in the limit of weak coupling and long interaction time, with negligible back-action of the detector on the field. We recall how the detailed balance condition, relating the detector's excitation and de-excitation rates, provides a notion of an effective temperature experienced by the detector, in general, dependent on the energy scale of the transitions. We also give an initial discussion about identifying regimes where the effective temperature might be lower than the ambient temperature. 
Finally, we present the response of a detector in inertial motion, in preparation for distinguishing effects due to acceleration from those due to speed.

\subsection{Field and detector}
We work in $(2+1)$-dimensional Minkowski spacetime, 
with a standard set of Minkowski coordinates $(t,x,y)$ and the metric 
$\dd s^2 = - \dd t^2 + \dd x^2 + \dd y^2$. 
In this spacetime we consider a quantised real scalar field $\phi$, 
with a dispersion relation that is isotropic in $(x,y)$ and subject to 
the mild monotonicity conditions specified in \Sref{sec:response-generaldispersion}, 
but otherwise arbitrary; 
in particular, we do not assume the dispersion relation to be Lorentz invariant. 
We denote by $\mathcal{H}_\phi$ the standard Fock space in which the positive frequencies are defined with respect to the timelike Killing vector~$\partial_t$. 

We assume that the field has been prepared in a thermal state in inverse temperature $\beta>0$, 
where the notion of thermality is with respect to the time evolution generated by~$\partial_t$. 
We assume that the thermal state has a Wightman two-point function, denoted by  
\begin{equation}
\label{eqn: thermal wightman}
\widetilde{\WW}_\beta(\xx',\xx'')~=~\braket{\phi(\xx')\phi(\xx'')}_\beta\,, 
\end{equation}
possibly modulo infrared subtleties that we shall describe shortly. 
$\widetilde{\WW}_\beta$ is not invariant under Lorentz boosts, 
not even when the dispersion relation is Lorentz invariant, 
because of the role of $\partial_t$ in the construction of the state: 
a heat bath has a distinguished rest frame. 

We probe the field by a pointlike detector in uniform circular motion, on the worldline
\begin{equation}
\label{eqn: trajectory}
\xx(t) ~=~ \bigl(t,R\cos(\Omega t), R\sin(\Omega t)\bigr)\,, 
\end{equation}
where $R>0$ is the orbital radius and $\Omega>0$ is the angular velocity. 
The orbital speed is $v = R\Omega$. We assume that the worldline is timelike, $v<1$. 
We have parametrised the worldline by the Minkowski time $t$ because this 
will give us a detector response that is appropriate for describing an analogue spacetime system,  
where $t$ is the `lab time' with respect to which any frequencies will be measured \cite{Gooding,Bunney:2023ude,Biermann}. 
For a genuinely relativistic detector, 
whose microphysics operates according to the relativistic proper time, 
the worldline should be parametrised by the relativistic proper time. 

The detector's Hilbert space is $\mathcal{H}_{\mathrm{D}} \simeq \CC^2$, 
spanned by the orthonormal basis $\bigl\{\ket{0}, \ket{1}\bigr\}$. 
The detector's Hamiltonian~$H_{\mathrm{D}}$, 
generating dynamics with respect to the Minkowski time~$t$, 
acts on $\mathcal{H}_{\mathrm{D}}$ as $H_{\mathrm{D}}\ket{0}=0$ and $H_{\mathrm{D}}\ket{1}=\Ebar\ket{1}$, 
where $\Ebar\in\RR \setminus\{0\}$. 
The detector is hence a two-level system, with energy gap $\Eabs$: 
for $\Ebar>0$, $\ket{0}$ is the ground state and $\ket{1}$ is the excited state; 
for $\Ebar<0$, the roles are reversed. 
We have included in the symbol $\Ebar$ the overline to emphasise that this energy is defined with respect to the Minkowski time, and the notation is thus adapted to the analogue spacetime system, 
where $t$ is the distinguished `lab time'. 
The conversion to a relativistic detector is by $\Ebar = \gamma E$, 
where $E$ is the energy with respect to the detector's proper time and $\gamma = {(1-v^2)}^{-1/2}$, 
and, for transition rates, by including the overall factor~$1/\gamma$. 

The total Hilbert space is $\mathcal{H}_\phi\otimes\mathcal{H}_{\mathrm{D}}$. 

In the interaction picture, and continuing to define time evolution with respect to the Minkowski time, 
we take the interaction Hamiltonian to be 
\begin{equation}
\label{eqn: interaction hamiltonian}
H_{\mathrm{I}} ~=~ \lambda\chi(t) \left( \frac{\dd}{\dd t}\phi\bigl(\xx(t)\bigr) \right) \otimes\mu(t)\,,
\end{equation} 
where $\mu$ is the detector's monopole moment operator, $\chi$ is a real-valued switching function that specifies how the interaction is turned on and off, and $\lambda$ is a real-valued coupling constant. 
Working to first order in perturbation theory in~$\lambda$, the probability for the detector to transition from $\ket{0}$ to $\ket{1}$, regardless of the final state of the field, 
is \cite{Unruh,birrell}
\begin{align}
\label{eqn: probability}
\mathcal{P}~=~\lambda^2|\braket{1|\mu(0)|0}|^2 \, \FF_\chi(\Ebar,\beta)\,,
\end{align}
where $\FF_\chi$ is the \emph{response function\/}, given by 
\begin{align}
\FF_\chi(\Ebar,\beta)~=~\int_{-\infty}^\infty \dd t'\int_{-\infty}^\infty \dd t''\,\chi(t')\chi(t'')
\, \ee^{-\ii \Ebar (t'-t'')} \, \WW_\beta(t',t'')\,, 
\label{eqn: Fchi}
\end{align}
and 
\begin{equation}
\label{eqn: derivative thermal wightman}
\WW_\beta(t',t'')
~=~ 
\left\langle \frac{\dd}{\dd t'}\phi\bigl(\xx(t')\bigr)\frac{\dd}{\dd t''}\phi\bigl(\xx(t'')\bigr) \right\rangle_\beta \,.
\end{equation}
As the factors in front of $\FF_\chi$ in \eqref{eqn: probability} are constants, 
independent of $\beta$, $\Ebar$ and the trajectory, we may, with a traditional abuse of terminology, refer to $\FF_\chi$ as the probability. 

We refer to $\WW_\beta$ \eqref{eqn: derivative thermal wightman} 
as the \emph{derivative correlation function\/}. 
Had $H_{\mathrm{I}}$ \eqref{eqn: interaction hamiltonian} not included the time derivative, 
$\WW_\beta(t',t'')$ in $\FF_\chi$ \eqref{eqn: Fchi} would have been replaced by 
the pullback of the Wightman function $\widetilde{\WW}_\beta$ \eqref{eqn: thermal wightman} to the detector's worldline, 
\begin{equation}
\label{eqn: thermal nonder wightman}
\widetilde{\WW}_\beta \bigl(\xx(t'),\xx(t'')\bigr)
~=~
\bigl\langle \phi\bigl(\xx(t')\bigr)\phi\bigl(\xx(t'')\bigr) \bigr\rangle_\beta\,. 
\end{equation}
For a massless Klein-Gordon field, $\widetilde{\WW}_\beta(\xx',\xx'')$ is however infrared divergent~\cite{Takagi}. We shall see that including the time derivative in $H_{\mathrm{I}}$ makes the detector's response well defined even for the massless Klein-Gordon field. 

As the thermal state is stationary with respect to the Killing vector $\partial_t$ and isotropic in $(x,y)$, the Wightman function is invariant under time translations along the detector's trajectory, so that $\WW_\beta(t',t'') = \WW_\beta(t'-t'',0)$. 
Dividing by the total duration of the interaction, 
and letting the duration tend to infinity, 
$\FF_\chi$ reduces to the stationary response function, 
\begin{align}
    \FF(\Ebar,\beta) 
    ~=~\int_{-\infty}^\infty \dd t\,\ee^{-\ii \Ebar t} \, \WW_\beta(t,0)\,,
\label{eqn: new response}
\end{align}
which is interpreted as the transition probability per unit time. 
The subtleties in the infinite duration limit are discussed in \cite{Fewster}; in particular, the limit assumes the coupling to tend to zero sufficiently fast for the first order perturbative treatment to remain valid in the limit.

From now on we work with the stationary response function
$\FF$ \eqref{eqn: new response}, 
and we refer to it as the response function.

\subsection{Response function mode sum}\label{sec:response-generaldispersion}

As the field's dispersion relation is by assumption spatially isotropic, 
the field mode frequency with respect to $\partial_t$ can be written as $\omega(|\bm{k}|)$, where 
$\bm{k}$ is the spatial momentum and $\omega(K)$ is function of a non-negative argument, 
positive everywhere except possibly at $K=0$. 
We write $\omega'(K) = \frac{\dd}{\dd K}\omega(K)$, 
and we assume that $\omega'(K) >0$ for $K>0$. 
Finally, if $\omega(0)=0$, we assume that $\omega'(0)>0$. 

We show in \ref{app: response derivation} that the response function has the mode sum expression 
\begin{align}
\label{eqn: general response function}
    \mathcal{F}(\Ebar,\beta) &~=~\frac{\Ebar^2}{2}
    \! 
    \left( \sum_{m>(\Ebar+\omega(0))/\Omega}\frac{K^+_m}{\omega'(K^+_m)\omega(K^+_m)} \, \Bigl(1+n\bigl(\beta\omega(K^+_m)\bigr)\Bigr)
    J^2_m(RK_m^+)
    \right.
    \nonumber 
    \\
    &\hspace{7ex}
    \left. 
    + \sum_{m>(-\Ebar+\omega(0))/\Omega}\frac{K^-_m}{\omega'(K^-_m)\omega(K^-_m)} \, n\bigl(\beta\omega(K^-_m)\bigr)J^2_m(RK_m^-)
    \right) 
    \,,
\end{align}
where 
\begin{align}
n(x)~=~\frac{1}{\ee^{x}-1}\,, 
\label{eq:nfactor-def}
\end{align}
$J_m$ are the Bessel functions of the first kind~\cite{NIST}, 
and $K_m^\pm$ is defined for $m > (\pm\Ebar+\omega(0))/\Omega$, as the unique solution to 
\begin{equation}
    \omega(K)-m\Omega\pm\Ebar~=~0 \,. 
\end{equation}
The uniqueness of $K_m^\pm$ follows from the positivity of $\omega'(K)$, 
and the notation suppresses the $\Ebar$-dependence of~$K_m^\pm$. 
Note that $n$ \eqref{eq:nfactor-def} 
is the Planckian factor characteristic of a thermal distribution in a bosonic field. 

If $\omega(0)=0$, the factors $n\bigl(\beta\omega(K^\pm_m)\bigr)/\omega(K^\pm_m)$ have singularities, 
but, by the assumption $\omega'(0)>0$, these singularities are more than outweighed by the factors 
$K_m^\pm J^2_m(RK_m^\pm)$ for $m\ne0$, whereas the $m=0$ term is not singular because $\Ebar\ne0$ by assumption; 
$\mathcal{F}(\Ebar,\beta)$ is hence continuous in $\Ebar$, but it is not smooth. 
We shall comment on this in more explicitly with the massless Klein-Gordon field below. 

\subsection{Massless Klein-Gordon field}\label{sec:KG-field}
We now specialise to the massless Klein-Gordon field, 
for which $\omega(K)=K$, $\omega'(K)=1$ and 
$K^\pm_m = m\Omega\mp\Ebar$.
The response function \eqref{eqn: general response function} then simplifies to 
\begin{subequations}
\label{eqn: response split collected}
\begin{align}
    \FF(\Ebar,\beta)&~=~\FF_0(\Ebar)+\DFB(\Ebar)\,,
    \label{eqn: response split}\\ 
    \label{eqn: vacuum contribution}
    \FF_0(\Ebar)&~=~\frac{\Ebar^2}{2}\sum_{m>\Ebar/\Omega}J^2_{m}\bigl((m\Omega-\Ebar) R\bigr)\,,\\
    \DFB(\Ebar)&~=~\frac{\Ebar^2}{2}
    \left(\sum_{m>\Eabs/\Omega} n\bigl((m\Omega-\Eabs)\beta\bigr) J^2_{m}\bigl((m\Omega-\Eabs) R\bigr) 
    \right.
    \nonumber \\
    & \hspace{10ex}
    \left. +\sum_{m>-\Eabs/\Omega} n\bigl((m\Omega+\Eabs)\beta\bigr) J^2_{m}\bigl((m\Omega+\Eabs) R\bigr)\right)\,,
    \label{eqn: thermal contribution}
\end{align}
\end{subequations}
where $\FF_0$ is the vacuum contribution, independent of $\beta$, while $\DFB$ is the additional contribution due to the ambient temperature. 
Note that $\DFB(\Ebar)$ is even in~$\Ebar$, 
and we have written \eqref{eqn: thermal contribution} in a way that makes this manifest. 
Note also that both $\FF_0$ and $\DFB$ are manifestly positive. 

Recall that by assumption $\Ebar \ne 0$ and $0<v<1$, where $v = R\Omega$. 
It follows from the uniform asymptotic expansion 10.20.4 in \cite{NIST} 
that the sums in \eqref{eqn: vacuum contribution} and \eqref{eqn: thermal contribution} converge, and $\FF(\Ebar,\beta)$ is hence well defined. 
$\FF(\Ebar,\beta)$ is however not smooth in $\Ebar$ at integer values of $\Ebar/\Omega$, 
where new terms enter the sums: at $\Eabs/\Omega = n \in \{1,2,\ldots\}$, 
$\DFB(\Ebar)$ has a discontinuity in its $(2n-1)^{\text{th}}$ derivative and 
$\FF_0(\Ebar)$ has a discontinuity in its $(2n)^{\text{th}}$ derivative. 

$\FF_0$ has the integral representation 
\begin{align}
\FF_0(\Ebar) ~=~ \Ebar^2 \! 
\left( \frac{\gamma}{4} 
- \frac{1}{2\pi} 
\int_0^\infty dz \, \frac{\sin\!\left(2 (\Ebar/\Omega) z\right)}{\sqrt{z^2 - v^2 \sin^2 \! z}} 
\right)
\,, 
\label{eq:ff0-integral}
\end{align}
where $\gamma = {(1-v^2)}^{-1/2}$, 
as follows by translating (4.3) in \cite{Biermann} 
to our analogue spacetime conventions and to our derivative-coupled interaction. 
This representation will be useful for discussing some of the 
limits in \Sref{sec: limits}.

\subsection{Detailed balance temperature}

We now describe the 
\emph{detailed balance temperature\/}, 
an effective, energy-dependent notion of temperature, which 
we use to quantify the detector's response. 

\subsubsection{Context}$\phantom{xxx}$\newline 
To set the context, recall that in a local system in equilibrium with a thermal bath, 
the excitation and de-excitation probabilities satisfy Einstein's \emph{detailed balance\/} condition \cite{Einstein,terhaar-book}, 
\begin{align}
\frac{P_{\downarrow}(\Delta)}{P_{\uparrow}(\Delta)} 
~=~ 
\ee^{\Delta/T}
\,,
\label{eq:detaliedbalance-generalsetting}
\end{align}
where 
$P_{\uparrow}(\Delta)$ is the exitation probability, 
$P_{\downarrow}(\Delta)$ is the de-excitation probability, $\Delta>0$ is the energy difference between the two states under consideration, and $T$ is the temperature of the thermal bath. Solving \eqref{eq:detaliedbalance-generalsetting} for $T$ gives
\begin{equation}
T ~=~
\frac{\Delta}{\displaystyle \ln \! \left(\frac{P_{\downarrow}(\Delta)}{P_{\uparrow}(\Delta)}\right)}
\,, 
\label{eq:Tdetaliedbalance-generalsetting}
\end{equation} 
which gives an operational way to determine the bath's temperature in terms of the probabilities $P_{\uparrow}(\Delta)$ and $P_{\downarrow}(\Delta)$ that are observable in the local system. 
Note that while both $P_{\uparrow}(\Delta)$ and $P_{\downarrow}(\Delta)$ depend on~$\Delta$, 
the temperature $T$ in \eqref{eq:detaliedbalance-generalsetting} and \eqref{eq:Tdetaliedbalance-generalsetting} does not: the temperature $T$ sets the ratio of the excitation and de-exitation probabilities for \emph{all\/} energy gaps. This is the characteristic feature of a local system in equilibrium with a thermal bath. 

In the linear acceleration Unruh effect, the excitation and de-excitation probabilities of a localised detector obey the detailed balance condition \eqref{eq:detaliedbalance-generalsetting} 
in the triple limit of weak coupling, long interaction time and sharp spatial localisation \cite{Unruh,DeWitt:1980hx}: 
$\Delta$ is the energy gap defined with respect to the relativistic proper time, 
and $T$ given by \eqref{eq:Tdetaliedbalance-generalsetting} is the Unruh temperature, $a/(2\pi)$, where $a$ is the relativistic proper acceleration. 
The reason behind this phenomenon is that the Minkowski vacuum is a genuine thermal state in the Fock space adapted to the boost Killing vector whose one orbit the linearly-accelerated observer follows~\cite{Unruh}, 
and the detector consequently responds to this thermality by excitations and de-excitations that obey detailed balance. 
Subtleties in the sense of the triple limit of weak coupling, long interaction time and sharp spatial localisation are discussed in \cite{Fewster,Lin:2006jw,benatti-floreanini-2004,DeBievre:2006pys}. 

For non-linear uniform accelerations, by contrast, Minkowski vacuum does not have a similar description as a genuine thermal state in a Fock space adapted to the accelerated motion. 
While the excitation and de-excitation probabilities are affected by the acceleration, the ratio of these probabilities 
depends on the energy gap in a way that does not follow the detailed balance exponential law \eqref{eq:detaliedbalance-generalsetting}, describable by the single parameter~$T$. 
The non-linear uniformly accelerated motions therefore do not have a conventional notion of temperature, and the physical phenomenon behind the acceleration effect may be described as a combination of synchrotron radiation and the conventional Unruh effect, as reviewed in~\cite{Fulling:2014}. In particular, the quantity $T$ given by \eqref{eq:Tdetaliedbalance-generalsetting} depends on the gap~$\Delta$. 

The probabilities $P_{\uparrow}(\Delta)$ and $P_{\downarrow}(\Delta)$ 
are however observable quantities in the local quantum system, 
affected by the acceleration, even when the acceleration is not linear. 
For a given value of~$\Delta$, the excitation and de-excitation probabilities 
in the local system are related as if the system were in equilibrium with a thermal bath in the temperature given by~\eqref{eq:Tdetaliedbalance-generalsetting}. 
The $\Delta$-dependent quantity given by \eqref{eq:Tdetaliedbalance-generalsetting} provides thus a useful quantifier of the system's response to the acceleration at energy gap~$\Delta$, and hence an operational notion of an effective temperature at a given energy scale. 
We call this quantity the \emph{detailed balance temperature\/}. 

To summarise: for the non-linear uniform accelerations the detailed balance temperature does not arise from an underlying thermal bath, but it is a useful quantifier of the response of the local quantum system at a given energy scale. This is the sense in which we employ the detailed balance temperature in this paper.

\subsubsection{Uniform circular motion}$\phantom{xxx}$\newline 
Returning to our uniform circular motion setting, 
we define the detailed balance temperature by 
\begin{equation}
\label{eqn: detailed balance}
T_{\mathrm{DB}}~=~\frac{\Ebar}{\displaystyle \ln \! \left(\frac{\FF(-\Ebar,\beta)}{\FF(\Ebar,\beta)}\right)}\,, 
\end{equation} 
where we recall that $\Ebar$ is the transition energy with respect to the Minkowski time (rather than the relativistic proper time), and the notation suppresses that $T_{\mathrm{DB}}$ may a priori depend on all the parameters of the problem, including~$\Ebar$. While $\Ebar$ can have either sign, the right-hand side of \eqref{eqn: detailed balance} is invariant under $\Ebar \to -\Ebar$, so that 
$T_{\mathrm{DB}}$ depends on $\Ebar$ only through the gap magnitude~$\Eabs$. 
As positive (respectively negative) $\Ebar$ corresponds to an excitation (a de-excitation),
we see that $T_{\mathrm{DB}}$ \eqref{eqn: detailed balance} agrees with~\eqref{eq:Tdetaliedbalance-generalsetting}, but with respect to Minkowski time rather than relativistic proper time. 

We thus adopt the detailed balance temperature $T_{\mathrm{DB}}$ \eqref{eqn: detailed balance} as an effective temperature at the energy scale~$\Eabs$. The notions of heating and cooling used below will compare $T_{\mathrm{DB}}$ with the ambient temperature~$1/\beta$. 

\subsection{Cooling inequality}

We shall find in Sections \ref{sec:high-ambient-temp} and \ref{sec: numerical}
that there are regimes where the circular motion detailed balance temperature $T_{\mathrm{DB}}$ is \emph{lower\/} than the ambient temperature~$\beta^{-1}$. Here we make preliminary observations as to where such parameter regimes might be found.

The $T_{\mathrm{DB}}$ definition \eqref{eqn: detailed balance} may be rearranged as 
\begin{align}
\label{eqn: alt-detailed-balance}
\FF(\Ebar,\beta)\ee^{\Ebar/T_{\mathrm{DB}}} ~=~ \FF(-\Ebar,\beta) 
\,.
\end{align} 
Whilst \eqref{eqn: alt-detailed-balance} holds for either sign of~$\Ebar$, 
let us assume here $\Ebar >0$, for simplicity of the notation. 
As $\FF$ is by construction positive,
\eqref{eqn: alt-detailed-balance} then shows that
the condition for $T_{\mathrm{DB}}$ to be lower than $\beta^{-1}$ is 
\begin{equation}
\label{eqn: cooling def}
    \FF(\Ebar,\beta)\ee^{\beta\Ebar}~<~\FF(-\Ebar,\beta)\,.
\end{equation} 
Using \eqref{eqn: response split collected}, 
and the evenness of $\DFB(\Ebar)$, this becomes 
\begin{equation}
\label{eqn: circ cooling inequal}
    \FF_0(\Ebar)\ee^{\beta\Ebar}+\DFB(\Ebar)(\ee^{\beta\Ebar}-1)~<~\FF_0(-\Ebar)\,.
\end{equation}

In the low temperature limit, $\beta\to\infty$, with the other parameters fixed, 
the leftmost term in \eqref{eqn: circ cooling inequal} 
shows that \eqref{eqn: circ cooling inequal} cannot hold. 

In the high temperature limit, $\beta\to 0^+$, with the other parameters fixed, 
we shall see in \Sref{sec:high-ambient-temp} that the left-hand side of \eqref{eqn: circ cooling inequal} has a finite limit, and the possibility of satisfying \eqref{eqn: circ cooling inequal} arises. We shall return to this analytically in \Sref{sec:high-ambient-temp} and numerically in \Sref{sec: numerical}.

\subsection{Inertial motion response function}\label{sec:inertial-response}

In this subsection we record the response function of a detector that is in inertial motion but with a nonvanishing velocity with respect to the heath bath. 
We shall use this in the later sections 
to distinguish the acceleration contribution from the velocity contribution in the circular motion response. 

Specialising to the massless Klein-Gordon field, the inertial motion response function may be obtained from \eqref{eqn: new response} with 
\eqref{eq:modefunctions}
and \eqref{eqn: Wtherm} in a straightforward way, 
using identities $6.671.1$ and $6.671.2$ in~\cite{G+R}. The outcome is 
\begin{equation}
\label{eqn: linear response}
    \FFL(\Ebar,\beta)~=~\frac{\Ebar^2}{2}
    \!
    \left(
    \gamma \Theta(-\Ebar)
    +
    \frac{1}{\pi}\int_{\frac{\Eabs}{1+v}}^{\frac{\Eabs}{1-v}}\frac{\dd x}{\left(\ee^{\beta x}-1\right)\sqrt{{(vx)}^2-{\bigl(x-\Eabs\bigr)}^2}}
    \right)
    \,,
\end{equation}
where $v$ is the detector's velocity in the heat bath's rest frame, satisfying $0<v<1$, and $\gamma=(1-v^2)^{-1/2}$ is the Lorentz factor. 
The subscript ``Lin'' stands for ``linear'', emphasising that the inertial motion may be viewed as the $R\to\infty$ limit of the circular motion \eqref{eqn: trajectory} with fixed orbital speed~$v = R\Omega$; as a consistency check, we have verified that \eqref{eqn: linear response} may be obtained from \eqref{eqn: response split collected} in this limit, viewing the sum as the Riemann sum of an integral and using the asymptotic expansions of the Bessel functions~\cite{NIST}. 
The integrand in \eqref{eqn: linear response} is singular at the upper and lower limits, 
but these singularities are integrable and the integral is well defined. 

An alternative expression for $\FFL(\Ebar,\beta)$ is 
\begin{equation}
\label{eqn: flin}
    \FFL(\Ebar,\beta)
    ~=~
    \frac{\Ebar^2\gamma}{2}
    \! 
    \left( 
    \Theta(-\Ebar)+\frac{1}{\pi}\int_{-\frac{\pi}{2}}^{\frac{\pi}{2}}\frac{\dd \theta}{\ee^{\beta\Eabs\gamma^2(1 + v\sin\theta)}-1}
    \right)\,, 
\end{equation}
obtained from \eqref{eqn: linear response}
by the substitution $x=\gamma^2\Eabs(1+v\sin\theta)$. 
\eqref{eqn: flin} is more convenient for extracting some asymptotic limits and for numerical evaluation, as the integrand is nonsingular over the whole integration range. 

\section{Asymptotic regimes}\label{sec: limits}

In this section we find analytic expressions for the response and the detailed balance temperature for circular motion in the asymptotic regimes of high and low ambient temperature, small energy gap, 
small orbital radius with fixed speed, 
and near-sonic speed. We also give the corresponding results for inertial motion, including there also the regime of large energy gap. 
We demonstrate that for both circular motion and inertial motion there are regimes in which the detailed balance temperature is lower than the ambient temperature.

\subsection{High ambient temperature}\label{sec:high-ambient-temp}

Consider the high ambient temperature limit, $\beta\rightarrow0^+$, 
with $\Omega$, $R$ and $\Ebar$ fixed. 

By 24.2.1 in~\cite{NIST}, 
the Planckian factor $n(x)$ \eqref{eq:nfactor-def} has the small argument Laurent expansion 
\begin{align}
n(x) ~=~
\sum_{k=0}^{\infty} \frac{B_{k}}{k!} \, x^{k-1}
~=~
\frac{1}{x} 
- \frac12 
+ \frac{x}{12} + \cdots
\,,
\label{eq:bernoulli-generator}
\end{align}
convergent for $0<x<2\pi$, 
where $B_k$ are the Bernoulli numbers. 
The Bessel function factors in \eqref{eqn: thermal contribution}
have exponential decay at large~$m$, by 
10.20.4	in~\cite{NIST}. It follows, by a dominated convergence argument, that the asymptotic expansion of $\DFB(\Ebar,\beta)$ at $\beta\rightarrow0^+$ may be found from \eqref{eqn: thermal contribution} by using \eqref{eq:bernoulli-generator} under the sum over $m$ and reversing the order of the sums. 
The expansion proceeds in powers $\beta^p$ with $p=-1,0,1,3,5,\ldots$, 
and to the leading order we have 
\begin{align}
    \DFB(\Ebar)&~=~\frac{\Ebar^2}{2\beta}
    \left(\sum_{m>\Eabs/\Omega} \frac{J^2_{m}\bigl((m\Omega-\Eabs) R\bigr)}{m\Omega-\Eabs}
    \ +\sum_{m>-\Eabs/\Omega} \frac{J^2_{m}\bigl((m\Omega+\Eabs) R\bigr)}{m\Omega+\Eabs} \right)
    \ + \OO(1)\,. 
\label{eqn: high-ambient-thermal-contribution}
\end{align}
For the detailed balance temperature, 
\eqref{eqn: response split collected}, 
\eqref{eqn: detailed balance}, 
and 
\eqref{eqn: high-ambient-thermal-contribution}
give 
\begin{align}
T_{\mathrm{DB}} = 
\frac{1}{\beta} {\mathcal Q}(v,\Eabs/\Omega) 
\ \ + \OO(1)\,, 
\label{eqn: high-ambient-detailed-balance-temperature}
\end{align}
where 
\begin{align}
{\mathcal Q}(v,k) ~=~ 
\frac{\displaystyle 
\sum_{m>-k} \frac{J^2_{m}\bigl((m+k) v\bigr)}{m+k} 
\ + 
\sum_{m>k} \frac{J^2_{m}\bigl((m-k) v\bigr)}{m-k}}{\displaystyle 
\sum_{m>-k} J^2_{m}\bigl((m+k) v\bigr)
\ - 
\sum_{m>k} J^2_{m}\bigl((m-k) v\bigr) }
\,, 
\label{eq:Qdouble-def}
\end{align}
$k$ is assumed positive, and we recall that $0<v<1$. 

The function ${\mathcal Q}(v,k)$ \eqref{eq:Qdouble-def} is well defined. 
The sums over $m$ converge, by the exponential falloff seen from  
10.20.4	in~\cite{NIST}, and the denominator is positive, 
as is seen by writing the denominator as 
\begin{align}
\frac{2}{\pi} 
\int_0^\infty dz \, \frac{\sin\!\left(2k z\right)}{\sqrt{z^2 - v^2 \sin^2 \! z}} 
\,, 
\label{eq:Qdouble-den-integral}
\end{align}
using \eqref{eqn: vacuum contribution} and~\eqref{eq:ff0-integral}. 
The positivity of \eqref{eq:Qdouble-den-integral} follows by 
breaking the integral into a sum of integrals over the intervals $\frac{\pi p}{2k} < z < \frac{\pi (p+1)}{2k}$, $p = 0,1,2,\ldots$, 
combining each even $p$ interval with the next odd $p$ interval, 
noting that the combined integrand in each term is then positive because the denominator in \eqref{eq:Qdouble-den-integral} is a strictly increasing function of~$z$, 
and finally observing that these rearrangements are justified by the convergence of \eqref{eq:Qdouble-den-integral} as an improper Riemann integral. 

At small $v$ with fixed $k$, ${\mathcal Q}(v,k)$ has the asymptotic behaviour 
\begin{align}
{\mathcal Q}(v,k) ~=~ 
\begin{cases}
{\displaystyle \frac{1}{k} - \tfrac12 v^2 + \OO(v^4) }
& 
\text{for}\ 0< k < 1
\,,
\\[2ex]
{\displaystyle \frac{1}{k\gamma} \left( 1 + \OO(v^4) \right)}
& 
\text{for}\ 1 \le k
\,,
\end{cases}
\label{eq:Qdouble-smallv}
\end{align}
as can be verified by expanding the sums in \eqref{eq:Qdouble-def} in $v$ term by term; 
interchanging the sum and the expansion is justified because the falloff
10.20.4	in \cite{NIST} allows differentiating the sums with respect to $v$ term by term for $0<v<1$. 
It follows that for fixed $k\ge1$, ${\mathcal Q}(v,k) < 1$ 
for sufficiently small~$v$. 

At $v\to1$ with fixed~$k$, ${\mathcal Q}(v,k)$ tends to zero, decaying proportionally to $1/\ln\gamma$. 
To see this, we note that the numerator in \eqref{eq:Qdouble-def} remains bounded as $v\to1$, by 10.20.4	in~\cite{NIST}, 
while the denominator diverges, with the leading term 
$\frac{4 \sqrt{3}}{\pi} k \ln\gamma$, 
as is seen using the integral representation \eqref{eq:Qdouble-den-integral}
and the asymptotic expansion in Appendix E of~\cite{Biermann}. 

Collecting these observations about ${\mathcal Q}(v,k)$, we see that for sufficiently high ambient temperatures, $T_{\mathrm{DB}}$ is lower than the ambient temperature for any fixed $\Eabs$ and sufficiently large~$v$, and also for any fixed $\Eabs \ge \Omega$ and sufficiently small~$v$. 
We shall return to ${\mathcal Q}(v,k)$ numerically in \Sref{sec: numerical}. 

We note in passing that while the sums stemming from $\DFB$ do not appear to have elementary analytic expressions, 
$\DFB(\Ebar)$ admits an elementary analytic bound at the special value $\Ebar = \pm\Omega$, at the demarcation between the two asymptotic behaviours shown in~\eqref{eq:Qdouble-smallv}: 
from \eqref{eqn: thermal contribution} we have 
\begin{align}
    \DFB(\pm\Omega)
    &~<~
    \frac{\Omega^2}{2}
    n(v\beta/R)
    \sum_{m=1}^\infty 
    \left( J^2_{m+1}(mv) + J^2_{m-1}(mv) \right) 
    \nonumber
    \\
    &~=~
    \frac{\Omega^2}{2}
    \frac{n(v\beta/R)}{\sqrt{1-v^2}}
    \,,
\label{eq:DeltaF-analytic-bound}
\end{align}
and the bound is sharper at lower values of~$\beta$. 
The inequality in \eqref{eq:DeltaF-analytic-bound} 
follows by renaming the summation index and replacing the Planckian factor by its value at the lowest summand, 
and the equality follows from the identity 
\begin{equation}
    \sum_{m=1}^\infty \bigl( J^2_{m-1}(mv)+J^2_{m+1}(mv) \bigr)~=~\frac{1}{\sqrt{1-v^2}}\,,
\label{eq:besselsum-closed}
\end{equation} 
which we verify in \ref{app: Bessel proof}. 
We have not found this identity in the existing literature.

\subsection{Low ambient temperature}

Consider the low ambient temperature limit, $\beta\to\infty$, 
with $\Omega$, $R$ and $\Ebar$ fixed. 

The Planckian factor $n(x)$ \eqref{eq:nfactor-def} has the large argument expansion 
\begin{align}
n(x) ~=~ \sum_{k=1}^{\infty} \ee^{-kx}
\,,
\label{eq:n-largearg-expansion}
\end{align}
convergent for $x>0$. 
By the exponential decay of the Bessel function factors in~\eqref{eqn: thermal contribution}, it follows by a dominated convergence argument
that the asymptotic expansion of $\DFB(\Ebar,\beta)$ as $\beta\rightarrow\infty$ may be found from \eqref{eqn: thermal contribution} by using \eqref{eq:n-largearg-expansion} and rearranging the sums. 
We find 
\begin{align}
    \DFB(\Ebar)&~=~
    \frac{\Ebar^2}{2}
    \left( 
    \ee^{-\beta(m^+ \Omega-\Eabs)} \, J^2_{m^+}\bigl((m^+ \Omega-\Eabs) R\bigr)
    + 
    \ee^{-\beta(m^- \Omega+\Eabs)} \, J^2_{m^-}\bigl((m^- \Omega+\Eabs) R\bigr)
    \right)
    \nonumber\\
    & \hspace{4ex}
    + \OO \! \left( \ee^{-2\beta \min \left(m^+ \Omega-\Eabs , m^- \Omega+\Eabs \right)}
    \right)\,, 
    \label{eq:deltaF-lowtemperature-twoterms}
\end{align}
where
\begin{align}
m^\pm ~=~1+\left\lfloor\pm \Eabs/\Omega\right\rfloor\,,
\end{align}
$\lfloor \, \cdot \, \rfloor$ is the floor function, 
and the notation suppresses the $\Ebar$-dependence of~$m^{\pm}$. 
Note that since 
\begin{align}
0 < m^{\pm}\Omega \mp \Eabs \le \Omega
\,, 
\end{align}
\eqref{eq:deltaF-lowtemperature-twoterms} shows 
that $\DFB(\Ebar)$ has an exponential decay in~$\beta$. 

The dominant term of the exponential decay can be determined by analysing how the exponents in \eqref{eq:deltaF-lowtemperature-twoterms} depend on~$\Ebar$. There are three qualitatively different cases: 
\begin{enumerate}
\item 
For 
$0 < \bigl| \Eabs - n\Omega \bigr| < \tfrac12 \Omega$, $n \in \{0,1,2,\ldots\}$: 
\begin{align}
    \DFB(\Ebar)
    ~=~
    \tfrac12 \Ebar^2 \, \ee^{-\beta\left|\Eabs - n\Omega\right|} 
    \, 
    J^2_{n}\bigl((\Eabs - n\Omega)R\bigr)
    \ + \ 
    \OO \! \left( \ee^{-\beta \min \left(2\left|\Eabs - n\Omega\right| , \Omega - \left|\Eabs - n\Omega\right| \right)}
    \right)
    \,. 
\label{eq:Fdiff-as-generic}
\end{align}
\item 
For 
$\Eabs = (n+\tfrac12) \Omega$, $n \in \{0,1,2,\ldots\}$: 
\begin{align}
    \DFB\bigl((n+\tfrac12)\Omega\bigr)
    ~=~
    \tfrac12 {(n+\tfrac12)}^2 \Omega^2 \, \ee^{-\beta\Omega/2}
    \! 
    \left( 
    J^2_{n+1}(\Omega R/2)
    + 
    J^2_{n}(\Omega R/2)
    \right)
    \ + \ \OO \! \left( \ee^{-\beta \Omega} \right)\,. 
\label{eq:Fdiff-as-half-integers}
\end{align}
\item 
For 
$\Eabs = n \Omega$, $n \in \{1,2,3,\ldots\}$: 
\begin{align}
    \DFB(n\Omega)~=~\tfrac12 n^2 \Omega^2 \, \ee^{-\beta\Omega}
    \! 
    \left( 
    J^2_{n+1}(\Omega R)
    + 
    J^2_{n-1}(\Omega R)
    \right)
    \ + \ \OO \! \left( \ee^{-2\beta \Omega} \right)\,. 
\label{eq:Fdiff-as-integers}
\end{align}
\end{enumerate}
The coefficient in the exponent of the large $\beta$ falloff is hence discontinuous in $\Eabs$ at integer values of~$\Eabs/\Omega$, by~\eqref{eq:Fdiff-as-integers}. 
At half-integer values of~$\Eabs/\Omega$, the coefficient in the exponent is continuous in~$\Eabs$, but the overall magnitude is discontinuous, by~\eqref{eq:Fdiff-as-half-integers}. 

The detailed balance temperature differs from the zero ambient temperature value by a correction that is exponentially decaying in $\beta$, with the same leading exponent as in \eqref{eq:Fdiff-as-generic}--\eqref{eq:Fdiff-as-integers}.

\subsection{Small gap}\label{sec: small energy}

Consider the limit $\Ebar\to0$, with fixed $\Omega$, $R$ and~$\beta$. 

In $\DFB(\Ebar,\beta)$ \eqref{eqn: thermal contribution}, 
for $\Eabs < \Omega$ we have 
\begin{align}
    \DFB(\Ebar)&~=~\frac{\Ebar^2}{2}
    \biggl\{
    n(\Eabs\beta) J^2_{0}(\Eabs R)
    \nonumber \\
    &
    + 
    \sum_{m=1}^\infty 
    \Bigl[ n\bigl((m\Omega-\Eabs)\beta\bigr) J^2_{m}\bigl((m\Omega-\Eabs) R\bigr) + 
    n\bigl((m\Omega+\Eabs)\beta\bigr) J^2_{m}\bigl((m\Omega+\Eabs) R\bigr)
    \Bigr] \biggr\} 
    \nonumber \\
    &~=~
    \frac{\Eabs}{2\beta}
    + \biggl(
    - \frac14
    + \sum_{m=1}^\infty 
    n(m\Omega\beta) J^2_{m}(m\Omega R) 
    \biggr) \Eabs^2
    + \left(\frac{\beta}{24} - \frac{R^2}{4\beta}\right) \! \Eabs^3
    \ + \OO \bigl(\Eabs^4\bigr)
\,,
\label{eqn: thermal-at-smallgap}
\end{align}
using \eqref{eq:bernoulli-generator}, and expanding in $\Eabs$ under the sum, justified by the falloff seen from 10.20.4 in~\cite{NIST}. 

For $\FF_0(\Ebar)$, we have 
\begin{align}
\FF_0(\Ebar) &~=~ 
\frac{\Ebar^2\gamma}{4}
- 
\frac{\Ebar^2 \sgn(\Ebar)}{4} 
+ \frac{\Ebar^3}{\pi\Omega}
\int_0^\infty dz \left( 1 - \frac{z}{\sqrt{z^2 - v^2 \sin^2\!z}} \right) 
+ \OO \bigl(\Ebar^4\bigr)
\,, 
\label{eqn: vacuum-at-smallgap}
\end{align}
obtained by applying to \eqref{eq:ff0-integral} the method of Appendix B of \cite{Biermann} 
and proceeding to one order higher to verify 
that the error term is as shown in~\eqref{eqn: vacuum-at-smallgap}. 

For the detailed balance temperature, 
\eqref{eqn: response split collected}, 
\eqref{eqn: detailed balance}, 
\eqref{eqn: thermal-at-smallgap}
and
\eqref{eqn: vacuum-at-smallgap}
give 
\begin{align}
T_{\mathrm{DB}} &~=~ 
\frac{1}{\beta} 
+ \left[
\frac{\gamma-1}{2} 
+ 2 
\sum_{m=1}^\infty 
    n(m\Omega\beta) J^2_{m}(m\Omega R)
+ \frac{4}{\pi\Omega\beta}
\int_0^\infty dz \left( 1 - \frac{z}{\sqrt{z^2 - v^2 \sin^2\!z}} \right) 
\right] \! \Eabs
\nonumber\\
&\hspace{5ex}
+ \OO\bigl(\Eabs^2\bigr)
\,, 
\end{align}
which reduces to the ambient temperature $1/\beta$ as $\Ebar\to0$.

\subsection{Large gap}\label{sec: large energy}

Consider the limit $\Ebar\to\pm\infty$, with fixed $\Omega$, $R$ and~$\beta$. 

For $\FF_0(\Ebar)$, using the integral representation \eqref{eq:ff0-integral} and adapting the analysis of \cite{Biermann} to our conventions shows that $\FF_0(\Ebar)$ consists of the inertial motion contribution and a piece that is exponentially suppressed in~$\Eabs$. 

Estimating $\DFB(\Ebar)$ \eqref{eqn: thermal contribution} at $\Eabs\to\infty$ would require new techniques. 
We shall not pursue this estimate here.

\subsection{Small radius with fixed speed}\label{sec:small-R} 

Consider the limit $R\to0$ with fixed $v$, $\beta$ and $\Ebar$. 

In $\DFB(\Ebar,\beta)$ \eqref{eqn: thermal contribution}, 
writing $\Omega = v/R$ gives, for 
$R < v/\Eabs$, 
\begin{align}
    \DFB(\Ebar)&~=~\frac{\Ebar^2}{2}
    \biggl\{
    n(\Eabs\beta) J^2_{0}(\Eabs R)
    \nonumber \\
    &\hspace{5ex}
    + 
    \sum_{m=1}^\infty 
    \Bigl[ n(mv/R-\Eabs\beta) J^2_{m}(mv-\Eabs R) + 
    n(mv/R+\Eabs\beta) J^2_{m}(mv+\Eabs R)
    \Bigr] \biggr\} 
    \nonumber \\
    &~=~
    \frac{\Ebar^2 n(\Eabs\beta)}{2}
    + \OO \bigl(R^2\bigr)
\,. 
\label{eqn: thermal-at-smallradius}
\end{align}

For $\FF_0(\Ebar)$, we have 
\begin{align}
\FF_0(\Ebar) &~=~ 
\frac{\Ebar^2\gamma}{4}
- 
\frac{\Ebar^2 \sgn(\Ebar)}{4} 
+ \frac{\Ebar^3 R}{\pi v}
\int_0^\infty dz \left( 1 - \frac{z}{\sqrt{z^2 - v^2 \sin^2\!z}} \right) 
+ \OO \bigl(R^2\bigr)
\,, 
\label{eqn: vacuum-at-smallradius}
\end{align}
obtained from \eqref{eq:ff0-integral} 
by writing $\Omega = v/R$ and proceeding as with~\eqref{eqn: vacuum-at-smallgap}. 

For the detailed balance temperature, 
\eqref{eqn: response split collected}, 
\eqref{eqn: detailed balance}, 
\eqref{eqn: thermal-at-smallradius}
and
\eqref{eqn: vacuum-at-smallradius}
give 
\begin{align}
T_{\mathrm{DB}} &~=~ 
\frac{\Eabs}{\displaystyle \ln \! \left(\frac{\gamma + 1 + 2n(\Eabs\beta)}{\gamma - 1 + 2n(\Eabs\beta)}\right)}
\ \ + \OO (R)
\,. 
\end{align}

Note that both 
$\FF(\Ebar)$ and $T_{\mathrm{DB}}$ remain finite as $R\to0$, despite the diverging acceleration. 

\subsection{Large radius with fixed speed}\label{sec:large-R} 

Consider the limit $R\to\infty$ with fixed~$v$, $\beta$ and $\Ebar$. 
This is the limit of inertial motion with speed~$v$, as discussed in \Sref{sec:inertial-response}. 
The leading term in $\FF$ is $\FFL$~\eqref{eqn: linear response}. 
We have not pursued the subleading corrections.

\subsection{Machian limit}\label{sec:machian} 

Consider the limit $R \to 1/\Omega$, with fixed $\Omega$, $\beta$ and~$\Ebar$. 
In terms of $v = \Omega R$, this is the limit $v\to1$ with fixed $\Omega$, $\beta$ and~$\Ebar$. In the analogue spacetime interpretation, this is the Machian, or near-sonic, limit. 

In $\DFB(\Ebar,\beta)$ \eqref{eqn: thermal contribution}, 
the $n$-factors have an exponential large $m$ falloff, while the falloff of the 
Bessel function factors is exponential for $0<v<1$ and $m^{-2/3}$ for $v=1$,  
as seen from 10.20.4 in~\cite{NIST}. 
It follows that $\DFB(\Ebar,\beta)$ has a finite limit as $v\to1$, 
\begin{align}
\DFB(\Ebar,\beta) &~=~ 
    \frac{\Ebar^2}{2}
    \left(\sum_{m>\Eabs/\Omega} n\bigl(\beta(m\Omega-\Eabs)\bigr) J^2_{m}(m-\Eabs/\Omega)
    \right.
    \nonumber \\
    & \hspace{10ex}
    \left. 
    +\sum_{m>-\Eabs/\Omega} n\bigl(\beta(m\Omega+\Eabs)\bigr) J^2_{m}(m+\Eabs/\Omega) \right)
    \ + o(1)\,. 
    \label{eqn: thermal-at-vto1}
\end{align}

For $\FF_0(\Ebar)$, the integral representation \eqref{eq:ff0-integral} and the formulas in Appendix E of \cite{Biermann} show that 
\begin{align}
\FF_0(\Ebar) &~=~ 
\Ebar^2 \left[
\frac{\gamma}{4} 
+ \frac{\sqrt{3}\, \Ebar}{\pi\Omega}
\ln \! \left(\frac{\sqrt{3} \, \ee^{\gamma_E -1} \Eabs}{\gamma\Omega}\right)
- \frac{1}{2\pi} h(2\Ebar/\Omega) 
\right]
\ + \ 
o(1)
\,,
\label{eqn: vacuum-at-vto1}
\end{align}
where $\gamma_E$ is the Euler-Mascheroni constant and 
\begin{align}
h(x) ~=~  \int_0^\infty dz \, \frac{\sin(xz)}{z} 
\left( \frac{1}{\sqrt{1 - (\sin^2 \!  z)/z^2}} - \frac{\sqrt{3}}{z} \right) 
\,. 
\end{align}

For the detailed balance temperature, 
\eqref{eqn: response split collected}, 
\eqref{eqn: detailed balance}, 
\eqref{eqn: thermal-at-vto1}
and 
\eqref{eqn: vacuum-at-vto1}
give 
\begin{align}
T_{\mathrm{DB}} ~=~ 
\frac{\pi \Omega}{8 \sqrt3} 
\! \left(\frac{\gamma}{\ln\gamma}\right)
\! 
\left\{
1 + 
\frac{1}{\ln\gamma}
\! 
\left[
\ln \! \left( \frac{\sqrt{3} \, \ee^{\gamma_E -1} \Eabs}{\Omega}\right)
- \frac{\Omega}{2 \sqrt3 \, \Eabs}
h(2\Eabs/\Omega)
\right]
\ + o \! \left(\frac{1}{\ln\gamma}\right)
\right\}
\,,
\end{align}
where all the terms shown come from~$\FF_0$: the ambient temperature enters only 
beyond the terms shown. 

\subsection{Inertial motion}\label{sec:inertial-limits}

In this subsection we record the corresponding asymptotic regime results for inertial motion. 

For large/small $\beta$ and large/small~$\Eabs$, we use \eqref{eqn: flin} and note that the expansions depend on $\beta$ and $\Eabs$ only through the combination $\beta\Eabs$. 
At high ambient temperature and/or small gap, $\beta\Eabs \to0$, we find 
\begin{subequations}
\begin{align}
\label{eqn: flin-smallproduct}
    \FFL(\Ebar,\beta)&~=~
    \frac{\Ebar^2}{2} \!  \left( 
    \frac{1}{\beta\Eabs}
    - \frac{\gamma \sgn(\Ebar) }{2} 
    + \frac{\gamma^3 \beta\Eabs}{12}
+ \OO\bigl(\beta^2\Ebar^2\bigr)
\right) \,, 
\\
T_{\mathrm{DB}}^{\mathrm{Lin}} &~=~ 
\frac{1}{\gamma\beta}
\! 
\left[ 
1
+ \frac{(\gamma-1)\gamma^2\beta^2\Ebar^2}{12}
+ \OO\bigl(\beta^4\Ebar^4\bigr)
\right] 
\,, 
\label{eqn: Tlin-smallproduct}
\end{align}
\end{subequations}
using \eqref{eq:bernoulli-generator}. 
At low ambient temperature and/or large gap, $\beta\Eabs \to\infty$, we find 
\begin{subequations}
\begin{align}
\label{eqn: flin-largeproduct}
    \FFL(\Ebar,\beta)&~=~
    \frac{\Ebar^2}{2} \!  \left[ 
    \gamma \Theta(-\Ebar)
    +\frac{\ee^{-\beta\Eabs/(1 + v)}}{\sqrt{2\pi v \beta \Eabs}} 
    \ + \OO \! \left({(\beta\Eabs)}^{-3/2} \ee^{-\beta\Eabs/(1 + v)}\right)
    \right] 
    \,, 
\\
T_{\mathrm{DB}}^{\mathrm{Lin}} &~=~ 
\frac{(1+v)}{\beta}
\! 
\left[ 1
+ 
\frac{(1+v)\ln(\beta\Eabs)}{2\beta\Eabs}
+ \OO \! \left(\frac{1}{\beta\Eabs}\right)
\right] 
\,, 
\label{eqn: Tlin-largeproduct}
\end{align}
\end{subequations}
using \eqref{eq:n-largearg-expansion} 
and the properties 10.32.1 and 10.40.1 from \cite{NIST} of the modified Bessel function~$I_0$. 

For the Machian limit, $v\to1$, we start from~\eqref{eqn: linear response}, 
change variables by 
$\omega = |\overline E| \left( 1 + z^2 \right)/({1+v})$, 
use a dominated convergence argument to take the $v\to1$ limit under the integral, and use 25.12.11 in~\cite{NIST}. We find 
\begin{subequations}
\begin{align}
\label{eqn: flin-v-to-1}
\FFL(\Ebar,\beta)&~=~
\frac{\Ebar^2}{2}
\! 
\left(
\gamma \Theta(-\Ebar)
+ 
\frac{1}{\sqrt{2\pi \beta \Eabs}}
\PolyLog_{\frac12} \! \bigl(\ee^{-\frac12\beta \Eabs}\bigr)
\ + o(1)
\right) 
\,,
\\
T_{\mathrm{DB}}^{\mathrm{Lin}} &~=~ 
\frac{\Eabs}{\ln\gamma}
\! 
\left[
1 + \frac{1}{\ln\gamma}
\ln\! \left(\frac{\PolyLog_{\frac12} \! \bigl(\ee^{-\frac12\beta \Eabs}\bigr)}{\sqrt{2\pi \beta \Eabs}} \right)
\ + o \! \left(\frac{1}{\ln\gamma}\right)
\right]
\,, 
\label{eqn: Tlin-machian}
\end{align}
\end{subequations}
where $\PolyLog$ is the polylogarithm~\cite{NIST}. 

In the low ambient temperature and/or large gap regime, the detailed balance temperature is always higher than the ambient temperature, by~\eqref{eqn: Tlin-smallproduct}. 
However, in the high ambient temperature and/or small gap regime, 
the detailed balance temperature is always lower than the ambient temperature, by~\eqref{eqn: Tlin-largeproduct}, 
and similarly in the Machian limit, 
by~\eqref{eqn: Tlin-machian}.

\section{Numerical results}\label{sec: numerical}

In this section we present numerical evidence about the detailed balance temperature in inertial motion and circular motion, interpolating between the asymptotic regimes considered in \Sref{sec: limits}.

\subsection{Detailed balance temperature in inertial motion}

Consider first the inertial motion. 

\begin{figure}[t]
    \centering
    \includegraphics[scale=1.4]{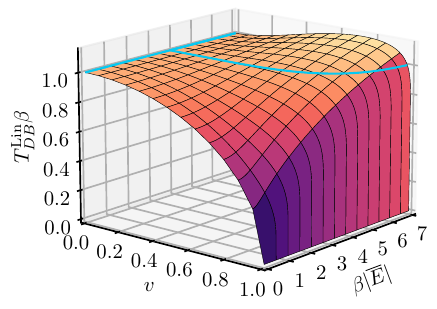}
    \caption{Inertial motion detailed balance temperature $T_{\mathrm{DB}}^{\mathrm{Lin}}$, plotted from \eqref{eqn: detailed balance} and~\eqref{eqn: flin}. The graph shows $\beta T_{\mathrm{DB}}^{\mathrm{Lin}}$ as a function of $v$ and~$\beta\Eabs$, which contains all the independent information. 
    The horizontal (blue) curve is at $\beta T_{\mathrm{DB}}^{\mathrm{Lin}} = 1$, at the boundary between a heating effect and a cooling effect. 
    The large $\beta\Eabs$ heating effect~\eqref{eqn: Tlin-largeproduct}, the small $\beta\Eabs$ cooling effect \eqref{eqn: Tlin-smallproduct} 
    and the large $v$ cooling effect \eqref{eqn: Tlin-machian} are evident in the plot; 
    note in particular the abrupt $1/\ln\gamma$ cooling effect \eqref{eqn: Tlin-machian} as $v\to1$. 
    The interpolation between the large $\beta\Eabs$ heating effect and the large $v$ cooling effect is showing in the region where both of these these quantities are large, with the horizontal (blue) curve 
    $\beta T_{\mathrm{DB}}^{\mathrm{Lin}} = 1$ 
    receding into the distance.}
    \label{fig:inertialfigure}
\end{figure}

All the independent information in 
$T_{\mathrm{DB}}^{\mathrm{Lin}}$ is obtained by expressing 
$\beta T_{\mathrm{DB}}^{\mathrm{Lin}}$ as a function of $v$ and~$\beta\Eabs$
by 
\eqref{eqn: detailed balance}
and~\eqref{eqn: flin}. 
A~plot is shown in \Fref{fig:inertialfigure}. The plot displays 
the interpolation between 
the large $\beta\Eabs$ heating effect~\eqref{eqn: Tlin-largeproduct}, 
the small $\beta\Eabs$ cooling effect \eqref{eqn: Tlin-smallproduct} 
and the large $v$ cooling effect~\eqref{eqn: Tlin-machian}. 

\subsection{Detailed balance temperature in circular motion}

Consider now the circular motion. 

\begin{figure}[p]
    \centering
    \includegraphics[scale=1.22]{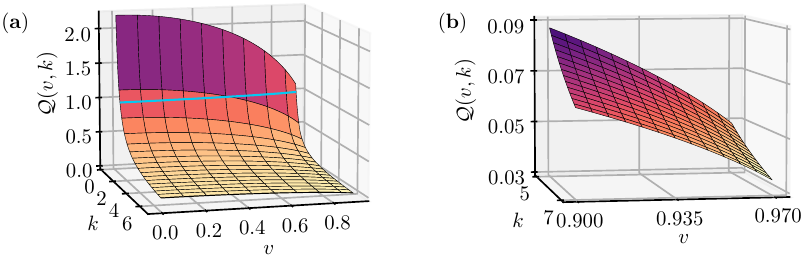}
    \caption{Plots of the function ${\mathcal Q}(v,k)$~\eqref{eq:Qdouble-def}, which determines the high ambient temperature asymptotics of $T_{\mathrm{DB}}$ by~\eqref{eqn: high-ambient-detailed-balance-temperature}. 
    Part (a) has $0.45 \le k \le 7$ and $0< v \le 0.9$, 
    and the horizontal (blue) curve is at ${\mathcal Q}(v,k) = 1$, 
    at the boundary between a heating effect and a cooling effect at high ambient temperature. Part (b) has $5 \le k \le 7$ and 
    $0.9 \le v \le 0.97$, showing incipient evidence of the $1/\ln\gamma$ falloff at $v\to 1$; closer to $v=1$ the numerics becomes slow because of the sums in the denominator in~\eqref{eq:Qdouble-def}. 
    If the pattern shown in the plots continues beyond the plotted range, there is a high ambient temperature cooling effect for all $\Eabs/\Omega\ge1$, regardless of~$v$.}
    \label{fig:Qfigure}
\end{figure}

As a preliminary, recall from \eqref{eqn: high-ambient-detailed-balance-temperature} that the high ambient temperature asymptotics of $T_{\mathrm{DB}}$ is determined by the function ${\mathcal Q}(v,k)$~\eqref{eq:Qdouble-def}, with $k= \Eabs/\Omega$. 
It was found in \Sref{sec:high-ambient-temp} that in this limit, 
there is a cooling effect for any fixed $\Eabs/\Omega$ and sufficiently large $v$, and also for any fixed $\Eabs/\Omega\ge1$ and sufficiently small~$v$. 
The plot of ${\mathcal Q}(v,k)$ in 
\Fref{fig:Qfigure} 
confirms numerically these analytic findings, and indicates that there is a 
cooling effect when $\Eabs/\Omega \ge 1$ for any~$v$, 
assuming the patterns in the plotted range continue outside the plotted range. 

\begin{figure}[p]
    \centering
\includegraphics[width=0.98\textwidth]{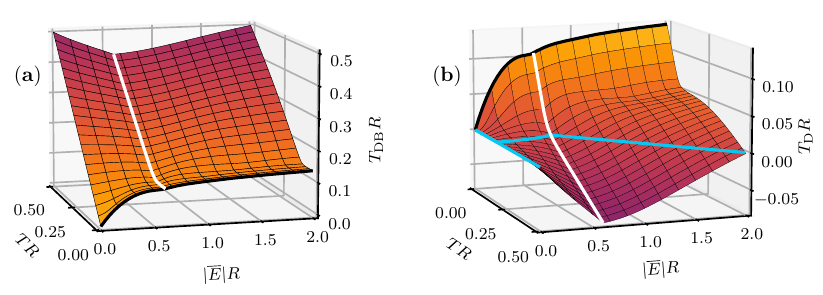}
    \caption{(a) Circular motion detailed balance temperature as a function of the ambient temperature and the energy gap, at fixed $v=0.6$. 
    The orbital radius $R$ enters only in that it sets the scale of the axes: 
    the vertical axis is $T_{\mathrm{DB}}R$, the ambient temperature horizontal axis is $TR$ where $T=1/\beta$, and the energy gap horizontal gap is $\Eabs R$. 
    The white curve marks the discontinuity in the first derivative at $\Eabs R=v$, 
    coming from the discontinuity in the first derivative of $\FF(\Ebar,\beta)$ \eqref{eqn: response split collected} at $\Eabs =\Omega = v/R$.\\
    (b) As in (a), but showing the difference of the detailed balance temperature and the ambient temperature: the vertical axis is $T_{\mathrm{D}}R$, where $T_{\mathrm{D}} = T_{\mathrm{DB}} - T$. 
    Note that the 
    horizontal $TR$ axes in (a) and (b) increase in opposite directions, 
    for the benefit of the visual perspective. The horizontal (blue) curve is at $T_{\mathrm{D}}R=0$, at the boundary between a heating effect and a cooling effect, and the white curve at $\Eabs R = v$ is at the discontinuity in the first derivative, as in~(a). A cooling effect near $\Eabs R = v$ sets in at moderate ambient temperature, from where it extends to high ambient temperature for $\Eabs R \ge v$.}
    \label{fig:DBfigure}
\end{figure}

Returning to finite ambient temperature, we address the interpolation between 
the asymptotic regimes analysed in \Sref{sec: limits} by plotting in 
\Fref{fig:DBfigure}(a) the detailed balance temperature as a function of the ambient temperature and the energy gap, 
for fixed $v = 0.6$ and a fixed orbital radius, 
and in \Fref{fig:DBfigure}(b) the difference of the detailed balance temperature and the ambient temperature. 
The orbital radius $R$ enters the plots only in that it sets the scales of the axes, 
and the range of the variables is chosen to cover the main transitional region of interest and to indicate the onset of asymptotics as numerical efficacy allows. 

The most prominent overall feature in \Fref{fig:DBfigure}(a) is that the ambient temperature dominates when the ambient temperature is high, as was to be expected, and as is consistent with the analytic estimates of \Sref{sec: limits}. 
The key information in \Fref{fig:DBfigure} 
is how the interpolation between the heating and cooling effects due to motion occurs: 
while we know from \Sref{sec: limits} and \Fref{fig:Qfigure} that a high ambient temperature cooling occurs for $\Eabs\ge\Omega$, 
\Fref{fig:DBfigure} shows
that this cooling effect occurs already for moderate 
ambient temperature if $\Eabs \approx \Omega$.

\section{Acceleration versus speed in circular motion}\label{sec:accel-versus-speed}

In this section we ask how much of the circular motion effect on the detector's response can be attributed to the detector's speed and how much to the acceleration. 

\subsection{Acceleration quantifiers}

A primary quantity indicating the significance of acceleration 
is the 
difference of the circular motion and inertial motion transition rates at the same speed. 
We quantify this difference by the ratio 
\begin{equation}
\label{eqn: normalised difference}
{\mathcal N}_v(\Ebar\beta,R/\beta)
~=~ 
\frac{\FF(\Ebar,\beta)-\FFL(\Ebar,\beta)}{\FFL(\Ebar,\beta)}
\,,
\end{equation}
where the notation on the right-hand side suppresses that 
$v$ in $\FFL$ is taken equal to $R\Omega$ in~$\FF$, and the notation on the left-hand side makes explicit that ${\mathcal N}_v$ depends on the parameters only through the dimensionless triple $(v,\Ebar\beta,R/\beta)$, 
as seen from \eqref{eqn: response split collected} and~\eqref{eqn: flin}. 
In words, ${\mathcal N}_v$ is the relative excess transition rate due to the acceleration at a given speed. 
Note that ${\mathcal N}_v$ may a priori be positive or negative. 
The acceleration contribution is insignificant iff $|{\mathcal N}_v| \ll 1$. 

A secondary quantity indicating the significance of acceleration is the ratio of the circular motion detailed balance temperature to the inertial motion detailed balance temperature, $T_{\mathrm{DB}}/T_{\mathrm{DB}}^{\mathrm{Lin}}$, 
at the same speed. 
Like ${\mathcal N}_v$, $T_{\mathrm{DB}}/T_{\mathrm{DB}}^{\mathrm{Lin}}$
depends on the parameters only through the dimensionless triple $(v,\Ebar\beta,R/\beta)$. 
Where is $T_{\mathrm{DB}}/T_{\mathrm{DB}}^{\mathrm{Lin}}$ approximately unity, and where is it significantly different from unity? 

\subsection{Asymptotic regimes}

We consider analytically four asymptotic regimes, 
using the results of \Sref{sec: limits}. 

First, in the limit $R\to\infty$ with fixed $v$, $\beta$ and~$\Ebar$, 
the trajectory becomes inertial, and the effects due to acceleration become insignificant by construction, as discussed in \Sref{sec:large-R}: 
we have ${\mathcal N}_v\to0$ and $T_{\mathrm{DB}}/T_{\mathrm{DB}}^{\mathrm{Lin}} \to 1$. 

Second, consider the limit $\Ebar\to0$ with fixed $v$, $R$ and~$\beta$. 
From Sections \ref{sec: small energy} and \ref{sec:inertial-limits} we obtain 
\begin{subequations}
\begin{align}
{\mathcal N}_v(\Ebar\beta,R/\beta)
&~=~ 
\beta\Eabs \left((\gamma-1)\Theta(\Ebar) 
+ 2 
\sum_{m=1}^\infty 
n(mv\beta/R) J^2_{m}(mv)\right) 
+ \OO\bigl(\Ebar^2\bigr)
\,,
\label{eq:Nv-small-Ebar}
\\
\frac{T_{\mathrm{DB}}}{T_{\mathrm{DB}}^{\mathrm{Lin}}} 
&~=~ 
\gamma + \OO(\Ebar)
\,.
\label{eq:Trat-small-Ebar}
\end{align}
\end{subequations}
The acceleration effect in the transition rate hence tends to zero linearly in $\Ebar$, but with different coefficients for excitations and de-excitations. The acceleration effect in the detailed balance temperature however remains nontrivial in this limit, increasing the temperature from $T_{\mathrm{DB}}^{\mathrm{Lin}}$ by the factor~$\gamma$. 

Third, consider the limit $R\to0$ with fixed~$v$, $\Ebar$ and~$\beta$. 
From Sections \ref{sec:small-R} and \ref{sec:inertial-limits} we find that 
both ${\mathcal N}_v$ and $T_{\mathrm{DB}}/T_{\mathrm{DB}}^{\mathrm{Lin}}$ have finite limits as $R\to0$, despite the diverging acceleration. The limits depend on the remaining variables only through the pair $(v,\Ebar\beta)$, 
but in a fairly complicated way: we find 
\begin{subequations}
\begin{align}
{\mathcal N}_v(\Ebar\beta,R/\beta)
&\xrightarrow[R\to0]{}
\frac{\bigl(\gamma - \sgn(\Ebar) + 2 n(\Eabs\beta)\bigr)}{\bigl(4/\Ebar^2\bigr) \FFL(\Ebar,\beta)} 
- 1
\nonumber
\\[1ex]
&\hspace{6ex} = 
\begin{cases}
{\displaystyle (\gamma-1) \Ebar\beta + \OO \bigl( {(\Ebar\beta)}^2 \bigr)}& \text{for}\ \Ebar\beta \to 0^+ \,, 
\\[1ex]
{\displaystyle - \frac{\gamma^3-1}{12} {(\Ebar\beta)}^2 + \OO \bigl( {(\Ebar\beta)}^3 \bigr)}& \text{for}\ \Ebar\beta \to 0^- \,, 
\\[2ex]
{\displaystyle \frac{\gamma-1}{2}  \sqrt{2\pi v \Ebar\beta} 
\, 
\ee^{\Ebar\beta/(1+v)} \! \left(1 
+ \OO \! \left( \frac{1}{\Ebar\beta}\right) \right)}& 
\text{for}\ \Ebar\beta \to \infty \,,
\\[1ex]
{\displaystyle - \frac12 \! \left(1 - \frac{1}{\gamma}\right)
+ \OO \! \left( \frac{\ee^{-\Eabs\beta/(1+v)}}{\sqrt{\Eabs\beta}} \right)}& \text{for}\ \Ebar\beta \to -\infty \,,
\end{cases}
\label{eq:Nv-smallR}
\end{align}
\begin{align}
\frac{T_{\mathrm{DB}}}{T_{\mathrm{DB}}^{\mathrm{Lin}}} 
&\xrightarrow[R\to0]{}
\frac{\displaystyle \ln \! \left(\frac{\FFL(-\Eabs,\beta)}{\FFL(\Eabs,\beta)}\right)}{\displaystyle \ln \! \left(\frac{\gamma + 1 + 2n(\Eabs\beta)}{\gamma - 1 + 2n(\Eabs\beta)}\right)}
\nonumber
\\[1ex]
&\hspace{6ex} = 
\begin{cases}
\gamma + \OO(\Eabs\beta) & \text{for}\ \Eabs\beta \to 0 \,, 
\\[1ex]
{\displaystyle \frac{\Eabs\beta}{(1+v) \ln \! \left(\frac{\gamma+1}{\gamma-1}\right)}
+ \OO\bigl(\ln(\Eabs\beta)\bigr)}& \text{for}\ \Eabs\beta \to \infty \,,
\end{cases}
\label{eq:Trat-smallR}
\end{align}
\end{subequations}
using the large and small $\Eabs\beta$ results from \Sref{sec:inertial-limits}. 
The $R\to0$ limit of ${\mathcal N}_v$ takes hence a wide range of values, from much less than unity to much larger than unity, depending on $\Ebar\beta$. 
The $R\to0$ limit of $T_{\mathrm{DB}}/T_{\mathrm{DB}}^{\mathrm{Lin}}$, by contrast, is larger than unity for both small and large~$\Eabs\beta$, and much larger than unity for large~$\Eabs\beta$. 

Fourth, consider the Machian limit, $v\to1$, with fixed $R$, $\Eabs$ and $\beta$. 
From Sections \ref{sec:machian} and \ref{sec:inertial-limits} we find 
\begin{subequations}
\begin{align}
{\mathcal N}_v(\Ebar\beta,R/\beta)
&~=~ 
\begin{cases}
{\displaystyle \frac{\sqrt{2\pi \Ebar\beta}}{2\PolyLog_{\frac12} \! \bigl(\ee^{-\frac12\Ebar\beta}\bigr)} \gamma  + o(\gamma)}& \text{for}\ \Ebar > 0 \,, 
\\[3ex]
{\displaystyle - \frac12 + \OO \! \left(\frac{\ln\gamma}{\gamma}\right) }& \text{for}\ \Ebar < 0 \,, 
\end{cases}
\\[1ex]
\frac{T_{\mathrm{DB}}}{T_{\mathrm{DB}}^{\mathrm{Lin}}} 
&~=~ 
\frac{\pi \gamma}{8 \sqrt3 \, \Eabs R} 
\left[
1 + \OO \! \left(\frac{1}{\ln\gamma}\right)
\right] 
\,,
\end{align}
\end{subequations}
showing a significant acceleration effect in 
$T_{\mathrm{DB}}/T_{\mathrm{DB}}^{\mathrm{Lin}}$, and in the excess excitation rate, but only a moderate suppression of the de-excitation rate.

\subsection{Numerical results}

We present numerical results for $v=0.6$, 
plotting ${\mathcal N}_v$ and $T_{\mathrm{DB}}/T_{\mathrm{DB}}^{\mathrm{Lin}}$ as a function of the independent dimensionless variables $\Ebar\beta$ and $R/\beta$. 

\begin{figure}[p]
    \centering
\includegraphics[scale=1.1]{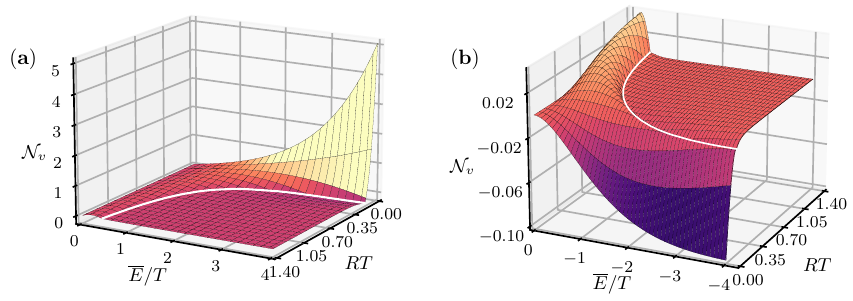}
    \caption{Relative excess transition rate ${\mathcal N}_v$ \eqref{eqn: normalised difference} due to acceleration, as a function of the circular trajectory radius $R$ and the excitation energy~$\Ebar$, at fixed $v=0.6$. 
    The ambient temperature $T = 1/\beta$ enters only in that it sets the scale of the horizontal axes, as shown. 
    Part (a) for excitations, $\Ebar>0$, and part (b) for de-excitations, $\Ebar<0$. The viewpoints are chosen
    for the benefit of the visual perspective. 
    The white curve is at $\Ebar R = v$, marking the discontinuity of the first derivative, as in \Fref{fig:DBfigure}. 
    At $\Eabs R \gg v$ we have 
    $|{\mathcal N}_v| \ll 1$, 
    but ${\mathcal N}_v$ starts to show nontrivial behaviour near $\Eabs R \approx v$, and the behaviour at $\Eabs R \ll v$ is consistent with the asymptotic estimates \eqref{eq:Nv-small-Ebar} and~\eqref{eq:Nv-smallR}. In particular, as $\Ebar/T\to0$, $N_v$ decays to zero linearly, by \eqref{eq:Nv-small-Ebar}; 
    in part (b) of the figure, this decay is mostly obscured by the hill in the plotted surface.}
    \label{fig:varying-radius}
\end{figure}

\begin{figure}[p]
    \centering
\includegraphics[scale=1.4]{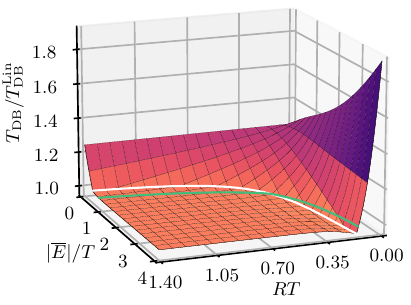}
    \caption{Ratio of the circular motion detailed balance temperature $T_{\mathrm{DB}}$ and the inertial motion detailed balance temperature $T_{\mathrm{DB}}^{\mathrm{Lin}}$, as a function of the circular trajectory radius $R$ and the excitation energy~$\Eabs$, at fixed $v=0.6$. 
    The ambient temperature $T = 1/\beta$ again enters only in the scales of the horizontal axes, and the white curve is at $\Eabs R = v$, at the discontinuity of the first derivative. 
    $T_{\mathrm{DB}}/T_{\mathrm{DB}}^{\mathrm{Lin}}$ is close to unity at $\Eabs R \gg v$, 
    but it starts to deviate significantly from unity near $\Eabs R \approx v$. The horizontal (green) curve is at $T_{\mathrm{DB}}/T_{\mathrm{DB}}^{\mathrm{Lin}}=1$. 
    The behaviour at $\Eabs R \ll v$ is consistent with the asymptotic estimates \eqref{eq:Trat-small-Ebar} and~\eqref{eq:Trat-smallR}; 
    in particular, at $\Ebar/T \to 0$, $T_{\mathrm{DB}}/T_{\mathrm{DB}}^{\mathrm{Lin}} \to \gamma = 5/4$, by~\eqref{eq:Trat-small-Ebar}.}
    \label{fig:temp-ratio-varying-radius}
\end{figure}

\Fref{fig:varying-radius} shows a plot of ${\mathcal N}_v$, 
both for $\Ebar>0$, corresponding to excitations, and for $\Ebar<0$, corresponding to de-excitations. 
In both cases, the plot indicates that $|{\mathcal N}_v|\ll1$ 
for $\Eabs R \gg v$, but significant deviations start near $\Eabs R \approx v$, and there are regions of positive ${\mathcal N}_v$ and regions of negative~${\mathcal N}_v$. 
We emphasise that the behaviour near $\Eabs R \approx v$ is independent of the ambient temperature, 
and this behaviour hence occurs even when the ambient temperature is so high that the ambient temperature dominates the overall magnitude of the detector's response. 
At $\Eabs R \ll v$, the behaviour in the plots is consistent with the asymptotic formulas given above. 

\Fref{fig:temp-ratio-varying-radius} shows a plot of $T_{\mathrm{DB}}/T_{\mathrm{DB}}^{\mathrm{Lin}}$, with the same parameter range as in \Fref{fig:varying-radius}. 
$T_{\mathrm{DB}}/T_{\mathrm{DB}}^{\mathrm{Lin}}$ is close to unity for $\Eabs R \gg v$, as had to happen by \Fref{fig:varying-radius}, but it starts to deviate significantly from unity near $\Eabs R \approx v$, where in some regions $T_{\mathrm{DB}}/T_{\mathrm{DB}}^{\mathrm{Lin}} \gg1$ but in some regions $T_{\mathrm{DB}}/T_{\mathrm{DB}}^{\mathrm{Lin}} < 1$. 
The behaviour at $\Eabs R \ll v$ is consistent with the asymptotic formulas given above. 

In summary, the combination of numerical an analytic evidence indicates that the acceleration has a negligible effect on the transition rate and on the detailed balance temperature for $\Eabs R \gg v$, but nontrivial behaviour due to the acceleration occurs at $\Eabs R \lesssim v$.

\section{Conclusions}\label{sec: conclusions}

Motivated by proposals to observe the analogue spacetime circular motion Unruh effect in condensed matter systems \cite{BEC1,Marino:2020uqj,Gooding,Bunney:2023ude}, 
we have analysed finite ambient temperature effects 
on a pointlike quantum system in uniform circular motion in $(2+1)$-dimensional Minkowski spacetime. 
We modelled the ambient quantum field by a massless Klein-Gordon field, prepared in a thermal state, and we modelled the pointlike quantum system by an UDW detector, coupled linearly to the time derivative of the field, where the time derivative is introduced to regularise an infrared divergence in the field's thermal Wightman function. We specified the detector's internal dynamics in terms of Minkowski time, instead of the detector's relativistic proper time, as this is expected to be more closely connected to quantities measured in an analogue spacetime experiment. 
We worked in the limit of weak interaction and long interaction time and neglected the detector's back-action on the field. 

We quantified the detector's response by an effective detailed balance temperature~$T_{\mathrm{DB}}$, 
computed from the detector's excitation and de-excitation rates by the detailed balance formula~\eqref{eqn: detailed balance}, 
and interpretable as an effective temperature within a limited energy band. 
We obtained analytic results in several asymptotic regimes and numerical results in the interpolating regimes. 

We found parameter regimes where $T_{\mathrm{DB}}$ is higher than the ambient temperature, so that the detector's motion causes a heating effect, 
and parameter regimes where $T_{\mathrm{DB}}$ is lower than the ambient temperature, so that the detector's motion causes a cooling effect. 
Comparing with an inertial detector, 
we found regimes where the heating/cooling is dominated by the detector's speed with respect to the heat bath, due to the Doppler effect, and the acceleration has only a minor role; conversely, we found regimes where the acceleration in the detector's motion is significant. 

While the interplay of the various regimes is subtle, one 
general feature is that a cooling effect is more likely to occur when the ambient temperature is high, as one might have expected on qualitative grounds. 
Another feature is that the detector's acceleration, rather than just the speed, 
is significant at energy gaps smaller than the orbital angular velocity, even when the ambient temperature is so high that the overall magnitude of the detector's transition rate is dominated by the ambient temperature. 

As a mathematical side product, we found an elementary expression for an infinite series \eqref{eq:besselsum-closed} involving squared Bessel functions. We have not encountered this identity in the existing literature. 

We recap that our motivation to consider a finite ambient temperature came from proposals to simulate an UDW detector on a circular orbit in a relativistic spacetime
by a condensed matter system, such as a phonons in a Bose-Einstein condensate or third sound waves in superfluid Helium, 
with a laser beam playing the role of the detector~\cite{Gooding,Bunney:2023ude}. 
These condensed matter systems simulate a Klein-Gordon field in a relativistic spacetime when probed at at sufficiently low frequencies, but they can never simulate a field state with a strictly zero temperature, and controlling finite ambient temperature effects is crucial. Our results chart the parameter space for these ambient temperature effects. 
As mentioned above, a specific upshot is that even when the ambient temperature is so high that the overall magnitude of the detector's response is dominated by the ambient temperature, as is likely be the case in superfluid Helium~\cite{Bunney:2023ude}, 
the acceleration effect can be distinguished from a Doppler shift effect by focusing on detector energy gaps smaller than the orbital angular velocity. This will inform the design of prospective experiments as analysed in~\cite{Bunney:2023ude}.

To summarise, 
an ambient temperature equips the circular motion Unruh effect 
with the characteristic of Robin Hood~\cite{Weinfurther-private-2023}: 
\textit{Where there is little, the Unruh effect gives; and where there is plenty, the Unruh effect takes.}

\ack

We thank the Quantum Simulators for Fundamental Physics discussion group, led by Silke Weinfurtner, for many helpful discussions, 
and Vitor S. Barroso for help with the numerics. 
We thank George Matsas for bringing the work in 
\cite{Costa:1994yx,Guimaraes:1998jf} to our attention. We thank the anonymous referees for helpful comments.
The work of JL was supported by United Kingdom Research and Innovation Science and Technology Facilities Council [grant number ST/S002227/1]. 
For the purpose of open access, the authors have applied a CC BY public copyright licence to any Author Accepted Manuscript version arising.

\appendix 

\section{Response function mode sum}\label{app: response derivation}

In this appendix we find the mode sum expression \eqref{eqn: general response function} for the response function. The notation and the assumptions are as in \Sref{sec:response-generaldispersion}. 

The positive Minkowski frequency mode functions of the massless Klein-Gordon field are 
\begin{align}
u_{\bm k}(\xx ) = 
\frac{1}{2\pi \sqrt{2 \omega}} \ee^{-\ii\omega t + \ii {\bm k} \cdot {\bm x}}
\,, 
\label{eq:modefunctions}
\end{align}
where ${\bm k}$ is the spatial momentum and the dispersion relation (suppressed in the notation) specifies $\omega$ 
as a function of~$|{\bm k}|$. 
The normalisation is $(u_{\bm k},u_{{\bm k}'}) = \delta_{{\bm k},{{\bm k}'}}$, where $(\cdot,\cdot)$ is the Klein-Gordon inner product
and $\delta$ is the Dirac delta. 
It follows as in \cite{Takagi} that the 
derivative correlation function $\mathcal{W}_\beta$ \eqref{eqn: derivative thermal wightman} has the mode sum expression 
\begin{align}
\label{eqn: Wtherm}
\mathcal{W}_\beta(t,t') &~=~ \int d^{2}\bm{k}
\left[ 
\bigl(1+n(\beta\omega)\bigr)\frac{\dd}{\dd t} u_{\bm{k}}\bigl(\xx (t)\bigr)\frac{\dd}{\dd t'}u^*_{\bm{k}}\bigl(\xx (t')\bigr) + n(\beta\omega) \frac{\dd}{\dd t}u^*_{\bm{k}}\bigl(\xx (t)\bigr)\frac{\dd}{\dd t'}u_{\bm{k}}\bigl(\xx (t')\bigr)
\right]\,,
\end{align}
where $\xx (t)$ is the circular motion worldline \eqref{eqn: trajectory}, 
$n$ is the Planckian factor \eqref{eq:nfactor-def} and the asterisk denotes complex conjugation. 

We substitute in \eqref{eqn: Wtherm} the mode functions \eqref{eq:modefunctions} and the trajectory \eqref{eqn: trajectory}, perform the differentiations, and change 
the integration variables from $(k_x,k_y)$ to $(k,l)$ by a suitable (time-dependent) rotation in the ${\bm k}$ plane, obtaining 
\begin{align}
\nonumber
\mathcal{W}_\beta(t,t')
& = 
\frac{1}{8\pi^2} 
\int\frac{\dd k \, \dd l}{\omega}
\left[\bigl(\omega-\Omega Rk\cos(\Omega s/2)\bigr)^2
- \Omega^2 R^2 l^2 \sin^2 (\Omega s/2)\right]\\
& \hspace{5ex}
\times \bigl[ 
\bigl(1+n(\beta\omega)\bigr) \ee^{-\ii\omega s+2\ii R k \sin(\Omega s/2)}
+ 
n(\beta\omega) \ee^{\ii\omega s-2\ii R k \sin(\Omega s/2)}
\bigr] 
\,,
\label{eq:Wbtt-1}
\end{align} 
where $s = t-t'$ and $\omega$ is now a function of $|\bm k| = \sqrt{k^2 + l^2}$. 
An equivalent expression is 
\begin{align}
\mathcal{W}_\beta(t,t')
& = 
\frac{1}{8\pi^2} 
\int\frac{\dd k \, \dd l}{\omega}
\biggl\{ 
\bigl(1+n(\beta\omega)\bigr)
\left[\bigl(\omega-\Omega Rk\cos(\Omega s/2)\bigr)^2
+ \tfrac{1}2 \ii \Omega^2 R k \sin(\Omega s/2)\right]
\nonumber
\\
& \hspace{18ex}
\times 
\ee^{-\ii\omega s+2\ii R k \sin(\Omega s/2)}
\nonumber
\\
& \hspace{5ex}
+ 
n(\beta\omega)
\left[\bigl(\omega-\Omega Rk\cos(\Omega s/2)\bigr)^2
- \tfrac{1}2 \ii \Omega^2 R k \sin(\Omega s/2)\right]
\ee^{\ii\omega s-2\ii R k \sin(\Omega s/2)}
\biggr\} 
\,, 
\label{eq:Wbtt-2}
\end{align} 
as can be seen by considering the difference of \eqref{eq:Wbtt-1} and \eqref{eq:Wbtt-2} 
and noting that the integral over the angle in the $(k,l)$ plane produces a sum of $J_0$, $J_0'$ and $J_0''$ that vanishes by Bessel's differential equation~\cite{NIST}.   
Note that the integrals in \eqref{eq:Wbtt-1} and \eqref{eq:Wbtt-2} are free of infrared divergences: when $\omega(0)=0$, this follows from the assumption $\omega'(0)>0$. 

We next use in \eqref{eq:Wbtt-2} the identity 
\begin{equation}
\label{eqn: jacobi anger}
    \ee^{\pm 2\ii Rk\sin(\Omega s/2)} ~=~ \sum_{n\in\ZZ}\ee^{\pm\ii n \Omega s/2} J_n(2Rk)\,, 
\end{equation}
which follows from 10.12.1 in~\cite{NIST}, 
and we regroup the sum so that all the $s$-dependence is in factors of the form 
$\ee^{ \mp \ii \omega s \pm\ii n \Omega s/2}$, multiplied by Bessel functions of order $n$, $n\pm1$ and $n\pm2$. 
We then convert the integral over $k$ and $l$ to polar coordinates by 
$(k,l)=(K\sin\theta,K\cos\theta)$, so that 
$\dd k \, \dd l = K \dd K \, \dd\theta$. 
The odd $n$ terms are odd in $\theta$ and vanish on integration over $\theta$. 
We relabel the even $n$ terms by $n=2m$ with $m\in\ZZ$, 
substitute in\eqref{eqn: new response}, and perform the integral over $t$ in terms of delta-functions, which collapse the integral over~$K$, with the outcome 
\begin{align}
    \mathcal{F}(\Ebar,\beta)~=~&\frac{1}{4\pi}\int_0^{2\pi} \dd\theta \, \Biggl\{ \, \sum_{m>(\Ebar+\omega(0))/\Omega}\frac{K^+_m}{\omega'(K^+_m)\omega(K^+_m)}
    \Bigl(1+n\bigl(\beta\omega(K^+)\bigr)\Bigr)\nonumber\\
    &\hspace{5ex}\times\biggl[\Bigl(\bigl(\omega(K_m^+)\bigl)^2+\tfrac{1}{2}\Omega^2 R^2 (K_m^{+})^2\sin^2\!\theta\Bigr)
    J_{2m}(2RK^+_m\sin\theta)\nonumber\\
    &\hspace{9ex}- \Omega RK^+_m \bigl(\omega(K^+_m)+\tfrac{1}{4}\Omega\bigr) \sin\theta J_{2m+1}(2RK^+_m\sin\theta) \nonumber\\
    &\hspace{9ex}- \Omega RK^+_m \bigl(\omega(K^+_m)-\tfrac{1}{4}\Omega \bigr) \sin\theta J_{2m-1}(2RK^+_m\sin\theta) \nonumber\\
    &\hspace{9ex}+\tfrac{1}{4}\Omega^2R^2 (K_m^{+})^2\sin^2\!\theta 
    \bigl( J_{2m+2}(2RK^+_m\sin\theta)+J_{2m-2}(2RK^+_m\sin\theta) \bigr)\biggr]\nonumber\\
&+\sum_{m>(-\Ebar+\omega(0))/\Omega}\frac{K^-_m}{\omega'(K^-_m)\omega(K^-_m)} \, n\bigl(\beta\omega(K^-)\bigr)
    \nonumber\\
&\hspace{5ex}\times\biggl[\Bigl(\bigl(\omega(K_m^-)\bigl)^2+\tfrac{1}{2}\Omega^2 R^2 (K_m^{-})^2\sin^2\!\theta\Bigr)
    J_{2m}(2RK^-_m\sin\theta)\nonumber\\
    &\hspace{9ex}- \Omega RK^-_m \bigl(\omega(K^-_m)+\tfrac{1}{4}\Omega\bigr) \sin\theta J_{2m+1}(2RK^-_m\sin\theta) \nonumber\\
    &\hspace{9ex}-\Omega RK^-_m \bigl(\omega(K^-_m)-\tfrac{1}{4}\Omega \bigr) \sin\theta J_{2m-1}(2RK^-_m\sin\theta) \nonumber\\
    &\hspace{9ex}+\tfrac{1}{4}\Omega^2R^2 (K_m^{-})^2\sin^2\!\theta 
    \bigl( J_{2m+2}(2RK^-_m\sin\theta)+J_{2m-2}(2RK^-_m\sin\theta) \bigr)\biggr]
\Biggr\}
\,, 
\label{eq:interm1}
\end{align} 
where $K_m^\pm$ are as defined in Section~\ref{sec:response-generaldispersion}, 
and the notation suppresses their $\Ebar$-dependence. 

The integrals in \eqref{eq:interm1} can be evaluated using the identities 
\begin{subequations}
\begin{align}
    \int_0^{2\pi}J_{2m}(2a\sin\theta)\,\dd\theta&~=~2\pi J^2_m(a)\,,\\
    \int_0^{2\pi}\sin\theta J_{2m\pm 1}(2a\sin\theta)\,\dd\theta&~=~2\pi J_{m\pm1}(a)J_m(a)\,,\\
    \int_0^{2\pi}\sin^2 \! \theta J_{2m}(2a\sin\theta)\,\dd\theta&~=~\pi \bigl( J^2_m(a)+J_{m-1}(a)J_{m+1}(a) \bigr)\,,
\end{align} 
\end{subequations} 
where $m\in\ZZ$; these identities follow from $6.681.1$ in \cite{G+R} by setting respectively $\mu=0$, $\mu=\frac{1}{2}$ and $\mu=1$. 
Further use of Bessel function identities 10.6.1 and 10.6.2 in \cite{NIST} 
then gives formula \eqref{eqn: general response function} in the main text. 

We end this appendix with two comments 
on the role of the time derivatives in~\eqref{eqn: general response function}. 

First, when $\omega(0)>0$, the result \eqref{eqn: general response function} can be obtained more shortly in the following way. 
The property $\omega(0)>0$ implies that the non-derivative correlation function $\widetilde{\WW}_\beta$ 
\eqref{eqn: thermal nonder wightman} is well defined, and  
\eqref{eqn: derivative thermal wightman} and \eqref{eqn: thermal nonder wightman} give 
$\WW_\beta(t',t'') = \partial_{t'} \partial_{t''} \widetilde{\WW}_\beta \bigl(\xx(t'),\xx(t'')\bigr)$. 
Using the time translation invariance of $\WW_\beta$ and $\widetilde{\WW}_\beta$, we then have 
\begin{align}
    \FF(\Ebar,\beta) 
    ~=~ \Ebar^2 \int_{-\infty}^\infty \dd t\,\ee^{-\ii \Ebar t} \, 
    \widetilde{\WW}_\beta \bigl(\xx(t),\xx(0)\bigr)\,, 
\label{eq:ebarsq-responserelation}
\end{align}
where the $\Ebar^2$ factor has come from integration by parts. 
The right-hand side of \eqref{eq:ebarsq-responserelation} is recognised as $\Ebar^2$ times the response function of a detector without a derivative in the coupling. 
We can now apply the methods of this appendix directly 
to the right-hand side of \eqref{eq:ebarsq-responserelation}, arriving at \eqref{eqn: general response function} through significantly fewer steps: 
the overall $\Ebar^2$ factor in \eqref{eqn: general response function} is exactly the overall $\Ebar^2$ factor in~\eqref{eq:ebarsq-responserelation}. 

Second, when $\omega(0)=0$, $\widetilde{\WW}_\beta$ is infrared divergent, and \eqref{eq:ebarsq-responserelation} as it stands is not well defined. 
However, if we ignore the infrared divergence in the mode sum expression for $\widetilde{\WW}_\beta$, 
and informally apply to \eqref{eq:ebarsq-responserelation} the 
integral interchange manipulations of this appendix, 
we arrive again at~\eqref{eqn: general response function}: 
the informal integral interchanges can be interpreted 
as a regularisation of the infrared divergence in 
the transition rate. 
This regularisation of the transition rate 
can be applied even when the detector's coupling does not involve a time derivative~\cite{Barman:2022utm}.

\section{Bessel function identity}\label{app: Bessel proof}

In this appendix we verify the identity 
\begin{equation}
\sum_{m=1}^\infty \bigl( J_{m-1}^2(mv)+J_{m+1}^2(mv) \bigr)
~=~
\frac{1}{\sqrt{1-v^2}}\,,
\end{equation}
where $0\le v<1$.

\begin{proof}
The case $v=0$ is trivial as the only nonvanishing term on the left-hand side is $J_0^2(0) = 1$. We henceforth assume $0<v<1$. 

Using the integral representation 10.9.26 in~\cite{NIST}, 
\begin{equation}
    J^2_n(z)~=~\frac{2}{\pi}\int_0^{\frac{\pi}{2}}J_{2n}(2z\cos\theta) \, \dd\theta\,, 
\end{equation}
we have 
\begin{align}
    \sum_{m=1}^\infty \bigl( J_{m-1}^2(mv)+J_{m+1}^2(mv) \bigr) &~=~ \frac{2}{\pi}\sum_{m=1}^\infty\int_0^{\frac{\pi}{2}}\left(J_{2m-2}(2mv\cos\theta)+J_{2m+2}(2mv\cos\theta)\right) \dd\theta 
    \nonumber \\
    & ~=~\frac{2}{\pi}\int_0^{\frac{\pi}{2}} \dd\theta 
     \sum_{m=1}^\infty\bigl(J_{2m-2}(2mv\cos\theta)+J_{2m+2}(2mv\cos\theta)\bigr) \,,
\label{eq:ident-int-1}
\end{align}
where the interchange of the sum and the integral is 
justified because the summands fall off exponentially in~$m$, 
uniformly in~$\theta$, 
as seen from 
10.20.4	in \cite{NIST}, 
recalling that $0 \le v\cos\theta \le v < 1$. 

For $0\le \theta < \pi/2$, we use Bessel function identities to rewrite the summands in \eqref{eq:ident-int-1} as 
\begin{equation}
    J_{2m-2}(2mv\cos\theta)+J_{2m+2}(2mv\cos\theta)~=~J_{2m}(2mv\cos\theta)\left(\frac{4}{v^2\cos^2 \theta}-2\right)-\frac{4}{v\cos\theta}\frac{J'_{2m}(2mv\cos\theta)}{2m}\,,
\end{equation}
and we then evaluate the sum over $m$ by the identities 
\begin{subequations}
\begin{align}
    \sum_{m=1}^\infty J_{2m}(2mt) &~=~ \frac{t^2}{2(1-t^2)}\,,\\
    \sum_{m=1}^\infty \frac{J'_{2m}(2mt)}{2m} &~=~ \frac{1}{2}\left(\sum_{k=1}^\infty \frac{J_k'(kt)}{k}-\sum_{k=1}\frac{J'_k(kt)}{k}(-1)^{k-1}\right)
    \nonumber\\
    &~=~\frac{1}{2}
    \left[
    \frac{1}{2}+\frac{t}{4}-\left(\frac{1}{2}-\frac{t}{4}\right)
    \right]
    \nonumber \\
    &~=~\frac{t}{4}\,,
\end{align} 
\end{subequations}
valid for $0 \le t < 1$, using   
8.517.3, 8.518.1 and 8.518.2 in~\cite{G+R}. 
Hence 
\begin{align}
    \sum_{m=1}^\infty \bigl(J_{m-1}^2(mv)+J_{m+1}^2(mv)\bigr)
    &~=~\frac{2}{\pi}\int_0^{\frac{\pi}{2}}
    \left[
    \left(\frac{4}{v^2\cos^2\theta}-2\right)\frac{v^2\cos^2\theta}{2(1-v^2\cos^2\theta)}-\frac{4}{v\cos\theta}\frac{v\cos\theta}{4}
    \right]
    \dd\theta 
    \nonumber\\
    &~=~\frac{2}{\pi}\int_0^{\frac{\pi}{2}}\frac{\dd\theta}{1-v^2\cos^2\theta} 
    \nonumber\\
    &~=~\frac{1}{\sqrt{1-v^2}}\,, 
\end{align}
where the last integral is elementary.
\end{proof}

\section*{References}
\bibliographystyle{iopart-num} 
\bibliography{bibliography}

\providecommand{\newblock}{}
\begin{thebibliography}{10}
\expandafter\ifx\csname url\endcsname\relax
  \def\url#1{{\tt #1}}\fi
\expandafter\ifx\csname urlprefix\endcsname\relax\def\urlprefix{URL }\fi
\providecommand{\eprint}[2][]{\url{#2}}

\bibitem{Fulling}
Fulling S~A 1973 {\em Phys. Rev. D\/} {\bf 7} 2850--2862

\bibitem{Davies1975}
Davies P~C~W 1975 {\em J. Phys. A: Math. Gen.\/} {\bf 8} 609--616

\bibitem{Unruh}
Unruh W~G 1976 {\em Phys. Rev. D\/} {\bf 14} 870--892

\bibitem{Fulling:2014}
Fulling S~A and Matsas G~E~A 2014 {\em Scholarpedia\/} {\bf 9} 31789

\bibitem{Hawking:1975vcx}
Hawking S~W 1975 {\em Commun. Math. Phys.\/} {\bf 43} 199--220 [Erratum:
  Commun. Math. Phys. 46, 206 (1976)]

\bibitem{Parker:1969au}
Parker L 1969 {\em Phys. Rev.\/} {\bf 183} 1057--1068

\bibitem{Mukhanov:2007zz}
Mukhanov V and Winitzki S 2007 {\em {Introduction to Quantum Effects in
  Gravity}\/} (Cambridge University Press)

\bibitem{Letaw}
Letaw J~R 1981 {\em Phys. Rev. D\/} {\bf 23} 1709--1714

\bibitem{Korsbakken:2004bv}
Korsbakken J~I and Leinaas J~M 2004 {\em Phys. Rev. D\/} {\bf 70} 084016
  (\textit{Preprint} \eprint{hep-th/0406080})

\bibitem{Good:2020hav}
Good M, Ju\'arez-Aubry B~A, Moustos D and Temirkhan M 2020 {\em JHEP\/} {\bf
  06} 059 (\textit{Preprint} \eprint{2004.08225})

\bibitem{LetawPfautsch}
Letaw J~R and Pfautsch J~D 1980 {\em Phys. Rev. D\/} {\bf 22} 1345--1351

\bibitem{Takagi}
Takagi S 1986 {\em Prog. Theor. Phys. Supp.\/} {\bf 88} 1--142

\bibitem{Doukas:2010wt}
Doukas J and Carson B 2010 {\em Phys. Rev. A\/} {\bf 81} 062320
  (\textit{Preprint} \eprint{1003.2201})

\bibitem{JIN201497}
Jin Y, Hu J and Yu H 2014 {\em Annals of Physics\/} {\bf 344} 97--104

\bibitem{Jin:2014spa}
Jin Y, Hu J and Yu H 2014 {\em Phys. Rev. A\/} {\bf 89} 064101
  (\textit{Preprint} \eprint{1406.5576})

\bibitem{BellLeinaas}
Bell J~S and Leinaas J~M 1983 {\em Nuclear Physics B\/} {\bf 212} 131--150

\bibitem{Bell:1986ir}
Bell J~S and Leinaas J~M 1987 {\em Nucl. Phys. B\/} {\bf 284} 488--508

\bibitem{Costa:1994yx}
Costa S~S and Matsas G~E~A 1995 {\em Phys. Rev. D\/} {\bf 52} 3466--3471
  (\textit{Preprint} \eprint{gr-qc/9412030})

\bibitem{Guimaraes:1998jf}
Guimaraes A~C~C, Matsas G~E~A and Vanzella D~A~T 1998 {\em Phys. Rev. D\/} {\bf
  57} 4461--4466 (\textit{Preprint} \eprint{hep-ph/9703309})

\bibitem{Leinaas:1998tu}
Leinaas J~M 1999 {Accelerated electrons and the Unruh effect} {\em {15th
  Advanced ICFA Beam Dynamics Workshop on Quantum Aspects of Beam Physics}\/}
  ed Chen P (World Scientific) pp 577--593 (\textit{Preprint}
  \eprint{hep-th/9804179})

\bibitem{Unruh:1998gq}
Unruh W~G 1998 {\em Phys. Rept.\/} {\bf 307} 163--171 (\textit{Preprint}
  \eprint{hep-th/9804158})

\bibitem{Lochan:2019osm}
Lochan K, Ulbricht H, Vinante A and Goyal S~K 2020 {\em Phys. Rev. Lett.\/}
  {\bf 125} 241301 (\textit{Preprint} \eprint{1909.09396})

\bibitem{BEC1}
Retzker A, Cirac J~I, Plenio M~B and Reznik B 2008 {\em \PRL\/} {\bf 101}
  110402 (\textit{Preprint} \eprint{0709.2425})

\bibitem{Marino:2020uqj}
Marino J, Menezes G and Carusotto I 2020 {\em Phys. Rev. Res.\/} {\bf 2} 042009
  (\textit{Preprint} \eprint{2001.08646})

\bibitem{Gooding}
Gooding C, Biermann S, Erne S, Louko J, Unruh W~G, Schmiedmayer J and
  Weinfurtner S 2020 {\em Phys. Rev. Lett.\/} {\bf 125} 213603
  (\textit{Preprint} \eprint{2007.07160})

\bibitem{Bunney:2023ude}
Bunney C~R~D, Biermann S, Barroso V~S, Geelmuyden A, Gooding C, Ithier G, Rojas
  X, Louko J and Weinfurtner S 2023  (\textit{Preprint} \eprint{2302.12023})

\bibitem{Unruh1981}
Unruh W~G 1981 {\em Phys. Rev. Lett.\/} {\bf 46} 1351--1353

\bibitem{Liberati}
Barcelo C, Liberati S and Visser M 2005 {\em Living Rev. Rel.\/} {\bf 8} 12
  (\textit{Preprint} \eprint{gr-qc/0505065})

\bibitem{HeliumUniverse}
Volovik G 2009 {\em The Universe in a Helium Droplet\/} International Series of
  Monographs on Physics (Oxford University Press)

\bibitem{Biermann}
Biermann S, Erne S, Gooding C, Louko J, Schmiedmayer J, Unruh W~G and
  Weinfurtner S 2020 {\em Phys. Rev D\/} {\bf 102} 085006 (\textit{Preprint}
  \eprint{2007.09523})

\bibitem{Muller:1995vk}
{M\"uller} D 1995  (\textit{Preprint} \eprint{gr-qc/9512038})

\bibitem{Hodgkinson}
Hodgkinson L, Louko J and Ottewill A~C 2014 {\em Phys. Rev. D\/} {\bf 89}
  104002 (\textit{Preprint} \eprint{1401.2667})

\bibitem{Barman:2022utm}
Barman S, Majhi B~R and Sriramkumar L 2022  (\textit{Preprint}
  \eprint{2205.01305})

\bibitem{DeWitt:1980hx}
DeWitt B~S 1979 {Quantum Gravity: The New Synthesis} {\em {General Relativity}:
  {An Einstein Centenary Survey}\/} ed Hawking S~W and Israel W (Cambridge:
  Cambridge University Press) pp 680--745

\bibitem{Fewster}
Fewster C~J, {Ju\'arez-Aubry} B~A and Louko J 2016 {\em Class. Quantum Grav.\/}
  {\bf 33} 165003 (\textit{Preprint} \eprint{1605.01316})

\bibitem{Lin:2006jw}
Lin S~Y and Hu B~L 2007 {\em Phys. Rev. D\/} {\bf 76} 064008 (\textit{Preprint}
  \eprint{gr-qc/0611062})

\bibitem{1991RSPSA.435..205R}
{Raine} D~J, {Sciama} D~W and {Grove} P~G 1991 {\em Proc. R. Soc. Lond. A\/}
  {\bf 435} 205--215

\bibitem{Raval:1995mb}
Raval A, Hu B~L and Anglin J 1996 {\em Phys. Rev. D\/} {\bf 53} 7003--7019
  (\textit{Preprint} \eprint{gr-qc/9510002})

\bibitem{Wang:2013lex}
Wang Q and Unruh W~G 2014 {\em Phys. Rev. D\/} {\bf 89} 085009
  (\textit{Preprint} \eprint{1312.4591})

\bibitem{Juarez-Aubry:2014jba}
Ju\'arez-Aubry B~A and Louko J 2014 {\em Class. Quant. Grav.\/} {\bf 31} 245007
  (\textit{Preprint} \eprint{1406.2574})

\bibitem{Juarez-Aubry:2018ofz}
Ju\'arez-Aubry B~A and Louko J 2018 {\em JHEP\/} {\bf 05} 140
  (\textit{Preprint} \eprint{1804.01228})

\bibitem{Juarez-Aubry:2021tae}
Ju\'arez-Aubry B~A and Louko J 2022 {\em AVS Quantum Sci.\/} {\bf 4} 013201
  (\textit{Preprint} \eprint{2109.14601})

\bibitem{Amelino-Camelia:2008aez}
Amelino-Camelia G 2013 {\em Living Rev. Rel.\/} {\bf 16} 5 (\textit{Preprint}
  \eprint{0806.0339})

\bibitem{Brenna:2015fga}
Brenna W~G, Mann R~B and Mart\'in-Mart\'inez E 2016 {\em Phys. Lett. B\/} {\bf
  757} 307--311 (\textit{Preprint} \eprint{1504.02468})

\bibitem{Liu:2016ihf}
Liu P~H and Lin F~L 2016 {\em JHEP\/} {\bf 07} 084 (\textit{Preprint}
  \eprint{1603.05136})

\bibitem{Garay:2016cpf}
Garay L~J, Mart\'in-Mart\'inez E and de~Ramon J 2016 {\em Phys. Rev. D\/} {\bf
  94} 104048 (\textit{Preprint} \eprint{1607.05287})

\bibitem{Li:2018xil}
Li T, Zhang B and You L 2018 {\em Phys. Rev. D\/} {\bf 97} 045005
  (\textit{Preprint} \eprint{1802.07886})

\bibitem{Henderson:2019uqo}
Henderson L~J, Hennigar R~A, Mann R~B, Smith A~R~H and Zhang J 2020 {\em Phys.
  Lett. B\/} {\bf 809} 135732 (\textit{Preprint} \eprint{1911.02977})

\bibitem{Pan:2020tzf}
Pan Y and Zhang B 2020 {\em Phys. Rev. A\/} {\bf 101} 062111 (\textit{Preprint}
  \eprint{2009.05179})

\bibitem{Pan:2021nka}
Pan Y and Zhang B 2021 {\em Phys. Rev. D\/} {\bf 104} 125014 (\textit{Preprint}
  \eprint{2112.01889})

\bibitem{DeSouzaCampos:2020ddx}
De~Souza~Campos L and Dappiaggi C 2021 {\em Phys. Lett. B\/} {\bf 816} 136198
  (\textit{Preprint} \eprint{2009.07201})

\bibitem{Barman:2021oum}
Barman S and Majhi B~R 2021 {\em JHEP\/} {\bf 03} 245 (\textit{Preprint}
  \eprint{2101.08186})

\bibitem{Zhou:2021nyv}
Zhou Y, Hu J and Yu H 2021 {\em JHEP\/} {\bf 09} 088 (\textit{Preprint}
  \eprint{2105.14735})

\bibitem{Robbins:2021ion}
Robbins M~P~G and Mann R~B 2022 {\em Phys. Rev. D\/} {\bf 106} 045018
  (\textit{Preprint} \eprint{2107.01648})

\bibitem{Chen:2021evr}
Chen Y, Hu J and Yu H 2022 {\em Phys. Rev. D\/} {\bf 105} 045013
  (\textit{Preprint} \eprint{2110.01780})

\bibitem{Wu:2022rmv}
Wu S~M, Zeng H~S and Liu T 2022 {\em New J. Phys.\/} {\bf 24} 073004
  (\textit{Preprint} \eprint{2207.01259})

\bibitem{birrell}
Birrell N~D and Davies P~C~W 1982 {\em Quantum Fields in Curved Space\/}
  Cambridge Monographs on Mathematical Physics (Cambridge University Press)

\bibitem{NIST}
{\it NIST Digital Library of Mathematical Functions} http://dlmf.nist.gov/,
  Release 1.1.9 of 2023-03-15 {F.~W.~J. Olver}, A.~B. {Olde Daalhuis}, D.~W.
  Lozier, B.~I. Schneider, R.~F. Boisvert, C.~W. Clark, B.~R. Miller, B.~V.
  Saunders, H.~S. Cohl, and M.~A. McClain, eds.

\bibitem{Einstein}
Einstein A 1917 {\em Phys. Z.\/} {\bf 18} 121--128

\bibitem{terhaar-book}
ter Haar D 1967 {\em The Old Quantum Theory\/} (Elsevier)

\bibitem{benatti-floreanini-2004}
Benatti F and Floreanini R 2004 {\em Phys. Rev. A\/} {\bf 70} 012112

\bibitem{DeBievre:2006pys}
De~Bi\`evre S and Merkli M 2006 {\em Class. Quant. Grav.\/} {\bf 23} 6525--6542
  (\textit{Preprint} \eprint{math-ph/0604023})

\bibitem{G+R}
Gradshteyn I~S and Ryzhik I~M 2007 {\em Table of {Integrals}, {Series}, and
  {Products}\/} seventh ed (Elsevier/Academic Press, Amsterdam)

\bibitem{Weinfurther-private-2023}
{Weinfurtner S} 2022 private communication

\end{thebibliography}

\end{document}